\documentclass[letterpaper,twocolumn,10pt]{article}
\usepackage[inchmargins, nocopyright]{sigmin}

\pdfpagewidth=\paperwidth
\pdfpageheight=\paperheight

\usepackage{times}
\usepackage{xspace}
\usepackage{color}
\usepackage[tight]{subfigure}
\usepackage{wrapfig}
\usepackage{multirow}
\usepackage{diagbox}
\usepackage{amssymb}
\usepackage[acronym,xindy,toc]{glossaries}
\usepackage{amsmath}
\usepackage[dvips]{graphicx}
\usepackage{listings}
\usepackage{url}
\usepackage{booktabs}
\usepackage[table]{xcolor}
\usepackage{./expl3}
\usepackage{collcell}
\usepackage{arydshln}
\newcommand{\bheading}[1]{{\vspace{4pt}\noindent{\textbf{{#1}.}}}} 

\newcommand{\bnm}{\begin{newmath}}
\newcommand{\enm}{\end{newmath}}

\newcommand{\bea}{\begin{eqnarray*}}
\newcommand{\eea}{\end{eqnarray*}}

\newcommand{\bne}{\begin{newequation}}
\newcommand{\ene}{\end{newequation}}

\newcommand{\docker}{Docker\xspace}

\newcommand{\heroku}{Heroku\xspace}
\newcommand{\openshift}{OpenShift\xspace}
\newcommand{\dotcloud}{DotCloud\xspace}

\newcommand{\appfog}{AppFog\xspace}

\newcommand{\azure}{Azure\xspace}
\newcommand{\elasticbeanstalk}{Elastic Beanstalk\xspace}
\newcommand{\engineyard}{Engine Yard\xspace}
\newcommand{\hpcloud}{HP Cloud\xspace}

\newcommand{\browse}{\texttt{browse}\xspace}
\newcommand{\logout}{\texttt{logout}\xspace}
\newcommand{\login}{\texttt{login}\xspace}
\newcommand{\post}{\texttt{post}\xspace}
\newcommand{\sending}{\texttt{send msg}\xspace}
\newcommand{\receiving}{\texttt{receive msg}\xspace}
\newcommand{\addfriend}{\texttt{add friend}\xspace}
\newcommand{\register}{\texttt{register}\xspace}
\newcommand{\update}{\texttt{update}\xspace}

\newcommand{\mysql}{\texttt{MySQL}\xspace}
\newcommand{\nginx}{\texttt{Nginx}\xspace}
\newcommand{\puma}{\texttt{Puma}\xspace}
\newcommand{\tornado}{\texttt{Tornado}\xspace}
\newcommand{\tomcat}{\texttt{Tomcat}\xspace}
\newcommand{\thin}{\texttt{Thin}\xspace}
\newcommand{\passenger}{\texttt{Passenger}\xspace}
\newcommand{\mongrel}{\texttt{Mongrel}\xspace}
\newcommand{\unicorn}{\texttt{Unicorn}\xspace}
\newcommand{\apache}{\texttt{Apache}\xspace}
\newcommand{\memcached}{\texttt{Memcached}\xspace}
\newcommand{\smem}{\texttt{smem}\xspace}
\newcommand{\autobench}{\texttt{autobench}\xspace}
\newcommand{\ssh}{\texttt{ssh}\xspace}
\newcommand{\php}{\texttt{PHP}\xspace}
\newcommand{\phpfpm}{\texttt{PHP-FPM}\xspace}
\newcommand{\phpinfo}{\texttt{phpinfo()}\xspace}
\newcommand{\ssl}{SSL\xspace}
\newcommand{\cgi}{cgi\xspace}
\newcommand{\cloudsuite}{CloudSuite\xspace}
\newcommand{\faban}{\texttt{Faban}\xspace}
\newcommand{\spin}{\texttt{Spin}\xspace}

\newcommand{\classNone}{\textsc{none}\xspace}
\newcommand{\classOne}{\textsc{one}\xspace}
\newcommand{\classFew}{\textsc{few}\xspace}
\newcommand{\classSome}{\textsc{some}\xspace}
\newcommand{\classLots}{\textsc{lots}\xspace}
\newcommand{\classMost}{\textsc{most}\xspace}
\newcommand{\classId}{\ensuremath{c}\xspace}
\newcommand{\classIdAlt}{\ensuremath{c'}\xspace}

\newenvironment{newmath}{\begin{displaymath}%
\setlength{\abovedisplayskip}{4pt}%
\setlength{\belowdisplayskip}{4pt}%
\setlength{\abovedisplayshortskip}{6pt}%
\setlength{\belowdisplayshortskip}{6pt} }{\end{displaymath}}

\newenvironment{newequation}{\begin{equation}%
\setlength{\abovedisplayskip}{4pt}%
\setlength{\belowdisplayskip}{4pt}%
\setlength{\abovedisplayshortskip}{6pt}%
\setlength{\belowdisplayshortskip}{6pt} }{\end{equation}}

\newcounter{packednmbr}

\newenvironment{packeditemize}{
\begin{list}{$\bullet$}{
\setlength{\labelwidth}{8pt}
\setlength{\itemsep}{0pt}
\setlength{\leftmargin}{\labelwidth}
\addtolength{\leftmargin}{\labelsep}
\setlength{\parindent}{0pt}
\setlength{\listparindent}{\parindent}
\setlength{\parsep}{0pt}
\setlength{\topsep}{3pt}}}{\end{list}}

\newcommand{\secref}[1]{\mbox{Sec.~\ref{#1}}\xspace}

\newcommand{\figref}[1]{\mbox{Fig.~\ref{#1}}}
\newcommand{\tabref}[1]{\mbox{Table~\ref{#1}}}
\newcommand{\appref}[1]{\mbox{App.~\ref{#1}}}
\newcommand{\eqnref}[1]{Eqn.~\ref{#1}\xspace}
\newcommand{\eqnsref}[2]{Eqns.~\ref{#1}--\ref{#2}\xspace}
\newcommand{\lstref}[1]{Line~\ref{#1}\xspace}
\newcommand{\ignore}[1]{}

\newcommand{\gbytes}{\ensuremath{\mathrm{GB}}\xspace}
\newcommand{\mbytes}{\ensuremath{\mathrm{MB}}\xspace}
\newcommand{\kbytes}{\ensuremath{\mathrm{KB}}\xspace}

\newcommand{\ghertz}{\ensuremath{\mathrm{GHz}}\xspace}

\newcommand{\secs}{\ensuremath{\mathrm{s}}\xspace}
\newcommand{\cycles}{\ensuremath{\mathrm{cycles}}\xspace}

\newcommand{\mbps}{\ensuremath{\mathrm{Mbps}}\xspace}

\newcommand{\flushreload}{\textsc{Flush-Reload}\xspace}
\newcommand{\flushflush}{\textsc{Flush-Flush}\xspace}
\newcommand{\Flush}{\textsc{Flush}\xspace}
\newcommand{\Reload}{\textsc{Reload}\xspace}
\newcommand{\primeprobe}{\textsc{Prime-Probe}\xspace}
\newcommand{\Prime}{\textsc{Prime}\xspace}
\newcommand{\Probe}{\textsc{Probe}\xspace}
\newcommand{\clflush}{\function{clflush}}

\newcommand{\function}[1]{\texttt{#1}\xspace}

\newcommand{\COA}{\gls{COA}\xspace}
\newcommand{\LRU}{NC\xspace}

\newcommand{\KSM}{\gls{KSM}\xspace}

\newcommand{\sysname}{\textsc{CacheBar}\xspace}
\newcommand{\sysnameLong}{Cache Barrier\xspace}
\newcommand{\lru}{cacheable queue\xspace}
\newcommand{\coa}{copy-on-access\xspace}

\newcommand{\unmapped}{\textsc{unmapped}\xspace}
\newcommand{\exclusive}{\textsc{exclusive}\xspace}
\newcommand{\shared}{\textsc{shared}\xspace}
\newcommand{\accessed}{\textsc{accessed}\xspace}
\newcommand{\accessedTimeout}{\ensuremath{\Delta_{\mathsf{accessed}}}\xspace}
\newcommand{\copyTimeout}{\ensuremath{\Delta_{\mathsf{copy}}}\xspace}

\newcommand{\ppage}{physical frame\xspace}

\newcommand{\pcounterincontainer}{\texttt{counter}\xspace}
\newcommand{\pcounterincontainers}[2]{\texttt{counter[{#1},{#2}]}\xspace}
\newcommand{\originalpagelist}{\texttt{original\_list}\xspace}
\newcommand{\copypagelist}{\texttt{copy\_list}\xspace}

\newcommand{\mapcount}{\texttt{\_mapcount}\xspace}

\newcommand{\owner}{\texttt{owner}\xspace}

\newcommand{\prob}[2]{\ensuremath{\mathbb{P}_{#1}\left({#2}\right)}\xspace}

\newcommand{\cprob}[4]{\ensuremath{\mathbb{P}_{#2}#1(}{#3}\ensuremath{\;#1|} \ifmmode{\;}\fi {#4}\ensuremath{#1)}\xspace}
\newcommand{\cexpv}[3]{\ensuremath{\mathbb{E}#1(}{#2}\ensuremath{\;#1|} \ifmmode{\;}\fi {#3}\ensuremath{#1)}\xspace}
\newcommand{\genericEvent}{\ensuremath{E}\xspace}
\newcommand{\victimDemand}{\ensuremath{d}\xspace}
\newcommand{\victimDemandAlt}{\ensuremath{d'}\xspace}
\newcommand{\attackerLineNmbr}{\ensuremath{\linesPerContainer{\attackerLabel}}\xspace}
\newcommand{\attackerLineNmbrRV}{\ensuremath{\linesPerContainerRV{\attackerLabel}}\xspace}
\newcommand{\victimLineNmbr}{\ensuremath{\linesPerContainer{\victimLabel}}\xspace}
\newcommand{\victimLineNmbrRV}{\ensuremath{\linesPerContainerRV{\victimLabel}}\xspace}
\newcommand{\cacheLineNmbr}{\ensuremath{w}\xspace}
\newcommand{\attackerNmbr}{\ensuremath{m}\xspace}
\newcommand{\containerNmbr}{\ensuremath{N}\xspace}
\newcommand{\containerIdx}{\ensuremath{i}\xspace}
\newcommand{\containerIdxAlt}{\ensuremath{i'}\xspace}
\newcommand{\linesPerContainer}[1]{\ensuremath{k_{#1}}\xspace}
\newcommand{\linesPerContainerRV}[1]{\ensuremath{K_{#1}}\xspace}
\newcommand{\attackerLabel}{\ensuremath{\mathsf{a}}\xspace}
\newcommand{\victimLabel}{\ensuremath{\mathsf{v}}\xspace}
\newcommand{\evictedNmbr}{\ensuremath{x}\xspace}
\newcommand{\evictedNmbrRV}{\ensuremath{X}\xspace}
\newcommand{\maxSecValue}{\ensuremath{\gamma}\xspace}
\newcommand{\maxPerfValue}{\ensuremath{\delta}\xspace}
\newcommand{\balanceSlack}{\ensuremath{\epsilon}\xspace}
\newcommand{\balanceValue}{\ensuremath{u}\xspace}
\newcommand{\physPageIdx}{\ensuremath{j}\xspace}
\newcommand{\classifiedRV}{\ensuremath{\mathsf{class}}\xspace}

\newacronym{DMcache}{DMcache}{Direct Mapped Cache}
\newacronym{PLcache}{PLCache}{Partition-Locked Cache}
\newacronym{RPcache}{RPCache}{Random-Permutation Cache}
\newacronym{LLC}{LLC}{Last Level Cache}
\newacronym{IPC}{IPC}{Inter-Process Communication}
\newacronym{LRU}{LRU}{Least-Recent-Used}
\newacronym{COA}{COA}{Copy-On-Access}
\newacronym{COW}{COW}{Copy-On-Write}
\newacronym{PTE}{PTE}{Page Table Entry}
\newacronym{KSM}{KSM}{Kernel Same-Page Merging}
\newacronym{MTRR}{MTRRs}{Memory Type range registers}
\newacronym{PAT}{PAT}{Page Attribute Table}
\newacronym{PCD}{PCD}{Page Cache Disable}
\newacronym{CR}{CR}{Control Registers}
\newacronym{UC}{UC}{Strong Uncacheable}
\newacronym{UCminus}{UC-}{Uncacheable}
\newacronym{WT}{WT}{write-through}
\newacronym{WC}{WC}{write-combining}
\newacronym{WB}{WB}{write-back}
\newacronym{WP}{WP}{write-protected}

\newcommand{\mcPages}[1]{\ensuremath{\mathtt{pages[{#1}]}}\xspace}
\newcommand{\mcVirt}{\ensuremath{\mathtt{virt}}\xspace}
\newcommand{\mcState}{\ensuremath{\mathtt{state}}\xspace}
\newcommand{\mcOwner}{\ensuremath{\mathtt{owner}}\xspace}
\newcommand{\mcNoOwner}{\ensuremath{\mathtt{none}}\xspace}

\ExplSyntaxOn
\fp_new:N \MinVal
\fp_new:N \MaxVal

\cs_new:Npn \intensity #1
   { \fp_eval:n
     { 0.5 + (0.5 * ({#1}-\MinVal)/(\MaxVal-\MinVal))
     }
   }
\ExplSyntaxOff

\newcommand{\ApplyGradientN}[1]{\cellcolor[gray]{\intensity{\MaxVal-{#1}}}{\parbox{2.4em}{\raggedleft {#1}}}}
\newcommand{\ApplyGradientR}[1]{\cellcolor[gray]{\intensity{#1}}{#1}}
\newcommand{\ApplyGradientT}[1]{\cellcolor[gray]{\intensity{#1}}{#1}}
\newcolumntype{N}{>{\collectcell\ApplyGradientN}c<{\endcollectcell}}
\newcolumntype{R}{>{\collectcell\ApplyGradientR}c<{\endcollectcell}}
\newcolumntype{T}{>{\collectcell\ApplyGradientT}c<{\endcollectcell}}

\begin{document}
%

\title{\Large \bf A Software Approach to Defeating Side Channels in Last-Level Caches}

\author{\begin{tabular}{ccc}
    Ziqiao Zhou & Michael K.\ Reiter & Yinqian Zhang\\[10pt]
    University of North Carolina & University of North Carolina & Ohio State University\\
    Chapel Hill, NC, USA & Chapel Hill, NC, USA & Columbus, OH, USA
    \end{tabular}}

\maketitle
\thispagestyle{empty}

\begin{abstract}
We present a software approach to mitigate access-driven side-channel
attacks that leverage last-level caches (LLCs) shared across cores to
leak information between security domains (e.g., tenants in a cloud).
Our approach dynamically manages physical memory pages shared between
security domains to disable sharing of LLC lines, thus preventing
``\flushreload'' side channels via LLCs.  It also manages cacheability
of memory pages to thwart cross-tenant ``\primeprobe'' attacks in
LLCs.  We have implemented our approach as a memory management
subsystem called \sysname within the Linux kernel to intervene on such
side channels across container boundaries, as containers are a common
method for enforcing tenant isolation in Platform-as-a-Service (PaaS)
clouds.  Through formal verification, principled analysis, and
empirical evaluation, we show that \sysname achieves strong security
with small performance overheads for PaaS workloads.
\end{abstract}

\section{Introduction}
\label{sec:intro}

An access-driven side channel is an attack by which an
attacker computation learns secret information about a victim
computation running on the same computer, not by violating the logical access
control implemented by the isolation software (typically an operating system
(OS) or virtual machine monitor (VMM)) but rather by observing the effects of
the victim's execution on microarchitectural components it shares with the
attacker's.  Overwhelmingly, the components most often used in these attacks
are CPU caches.  Early cache-based side channels capable of leaking fine-grained
information (e.g., cryptographic keys) across security boundaries used per-core
caches (e.g.,~\cite{tromer2010efficient,gullasch2011games,zhang2012cross}),
though the need for the attacker to frequently preempt the victim to observe its
effect on per-core caches renders these attacks relatively easy to mitigate in
software (e.g.,~\cite{zhang2013duppel,
varadarajan2014scheduler}).\footnote{Hyper-threading can enable the attacker to
observe the victim's effects on per-core caches without preempting it, if both
are simultaneously scheduled on the same core.  So, potentially adversarial
tenants are generally not scheduled together (or hyper-threading is disabled) in
cloud environments, for example.}  Of more concern are side channels via
last-level caches (LLCs) that are shared across
cores and, in particular, do not require preemption of the victim to extract
fine-grained information from it
(e.g.,~\cite{yarom2014flush,zhang2014cross,irazoqui2015shared,liu2015practical}).

Two varieties of LLC-based side channels capable of extracting
fine-grained information from a victim have been demonstrated.  The
first such attacks were of the \flushreload
variety~\cite{yarom2014flush,zhang2014cross}, which requires the
attacker to share a physical memory page with the victim---a common
situation in a modern OS, due to shared library, copy-on-write memory
management and memory deduplication mechanisms that aim for smaller
memory footprints.  The attacker first \Flush{es} a cache-line sized
chunk of the shared page out of the cache using processor-specific
instructions (e.g., \clflush in x86 processors) and later measures the
time to \Reload (or re-\Flush~\cite{gruss:2015:FF}) it to infer
whether this chunk was touched (and thus loaded to the shared cache
already) by the victim.  More recently, so-called \primeprobe attacks
have been demonstrated via
LLCs~\cite{irazoqui2015shared,liu2015practical}; these do not require
page sharing between the attacker and victim.  Rather, \primeprobe
attacks can be conducted when the two programs share the same CPU
cache sets. The attacker \Prime{s} the cache by loading its own memory
into certain cache sets. Later it \Probe{s} the cache by measuring the
time to load the same memory into the cache sets and inferring how
many cache lines in each cache set are absent due to conflicts with
the victim's execution.

In this paper we propose a software-only defense against these
LLC-based side-channel attacks, based on two seemingly straightforward
principles. First, to defeat \flushreload attacks, we propose a \coa
mechanism to manage physical pages shared across mutually distrusting
\textit{security domains} (i.e., processes,
containers\footnote{https://linuxcontainers.org/}, or VMs).
Specifically, temporally proximate accesses to the same physical page
by multiple security domains results in the page being copied so that
each domain has its own copy.  In this way, a victim's access to its
copy will be invisible to an attacker's \Reload in a \flushreload
attack. When accesses are sufficiently spaced in time, the copies can
be deduplicated to return the overall memory footprint to its original
size. Second, to defeat \primeprobe attacks, we design a mechanism to
manage the cacheability of memory pages so as to limit the number of
lines per cache set that an attacker may \Probe. In doing so, we
limit the visibility of the attacker into the victim's demand for
memory that maps to that cache set.  Of course, the challenge in these
defenses is in engineering them to be effective in both mitigating
LLC-based side-channels and supporting efficient execution of
computations.

To demonstrate these defenses and the tradeoffs between security and
efficiency that they offer, we detail their design and implementation
in a memory management subsystem called \sysname (short for
``\sysnameLong'') for the Linux kernel.  \sysname supports these
defenses for security domains represented as Linux
containers.  That is, \coa to defend against \flushreload attacks
makes copies of pages as needed to isolate temporally proximate
accesses to the same page from different containers.  Moreover, memory
cacheability is managed so that the processes in each container are
collectively limited in the number of lines per cache set they can
\Probe.  This implementation would thus be well-suited for use in
Platform-as-a-Service (PaaS) clouds that isolate cloud customers in
distinct containers; indeed, cross-container LLC-based side channels
have been demonstrated in such clouds in the
wild~\cite{zhang2014cross}.  Our security evaluations show that
\sysname mitigates cache-based side-channel attacks, and our
performance evaluation indicates that \sysname imposes very modest
overheads on PaaS workloads.

To summarize, we contribute:

\begin{packeditemize}

\item A novel copy-on-access mechanism to manage physical memory pages
  shared by distrusting tenants to prevent \flushreload side-channel
  attacks, and its formal verification using model checking.

\item A novel mechanism to dynamically maintain queues of cacheable
  memory pages so as to limit the cache lines a malicious tenant may
  access in \primeprobe attacks, and a principled derivation of its
  parameters to balance security and performance.
  
\item Implementation of both mechanisms in a mainstream Linux
  operating system kernel and an extensive security and performance
  evaluation for PaaS workloads.
\end{packeditemize}

\section{Related Work}
\label{sec:related}

Numerous proposals have sought to mitigate cache-based side channels
with low overhead through redesign of the cache hardware,
e.g.,~\cite{page2005partitioned, kong2008deconstructing,
  wang2008novel, keramidas2008non, Liu:2014:RFC}.  Unfortunately,
there is little evidence that mainstream CPU manufacturers will deploy
such defenses in the foreseeable future, and even if they did, it
would be years before these defenses permeated the installed computing
base.  Other proposals modify applications to better protect secrets
from side-channel attacks.  These solutions range from tools to limit
branching on sensitive data (e.g.,~\cite{coppens2009practical,
  crane2015thwarting}) to application-specific side-channel-free
implementations (e.g.,~\cite{konighofer2008fast}).  These techniques
can introduce substantial runtime overheads, however, and these
overheads tend to increase with the generality of the tool.

It is for this reason that we believe that systems-level (i.e., OS- or
VMM-level) defenses are the most plausible, general defense for
deployment in the foreseeable future.  With attention to cache-based
side-channels specifically, several works provide to each security
domain a limited number of designated pages that are never evicted
from the LLC (e.g.,~\cite{kim2012stealthmem,CATalyst}), thereby
rendering their contents immune to \primeprobe and \flushreload
attacks.  These approaches, however, require the application developer
to determine what data/instructions to protect and then to modify the
application to organize the sensitive content into the protected
pages; in contrast, \sysname seeks to protect applications
holistically and requires no application modifications.  \sysname also
differs in several design choices that free it from limitations of
prior approaches (e.g., the limitation of only one protected page per
core~\cite{kim2012stealthmem} or dependence on relatively recent,
Intel-specific cache optimizations~\cite{CATalyst}).  Other
systems-level solutions manage memory so as to partition the use of
the LLC by different security domains
(e.g.,~\cite{raj2009resource,shi2011limiting}), though these
approaches preclude memory-page and CPU-cache sharing entirely and
hence can underutilize these resources considerably.

LLC-based side channels are a particular instance of timing side
channels, and so defenses that seek to eliminate timing side channels
are also relevant to our problem.  Examples include fuzzing real-time
sources on the computer (e.g.,~\cite{vattikonda2011timers}), though
this impinges on legitimate uses of real time.  Since real-time
counters are not the only way to time memory
fetches~\cite{wray1991analysis}, other efforts have sought to
eliminate side-channel risks more holistically via altering the CPU
scheduler (e.g.,~\cite{stefan2013eliminating,li2014stopwatch}) and
managing how tenants co-locate
(e.g.,~\cite{li2012improving,zhang2012incentive,han2013security,azar2014colocation,li2014stopwatch}).
In contrast, here we focus specifically on LLC-based side channels
(vs.\ a larger subset of timing side-channels)---which again are
arguably the most potent known side-channel
vectors~\cite{yarom2014flush,zhang2014cross,irazoqui2015shared,liu2015practical}---and
restrict our modifications to the memory management subsystem.

\section{Copy-On-Access For \flushreload Defense}
\label{sec:coa}

The \flushreload attack is a highly effective LLC-based side channel
that was used, e.g., by Zhang et al.~\cite{zhang2014cross} to mount
fine-grained side-channel attacks in commercial PaaS clouds.  It
leverages physical memory pages shared between an attacker and victim
security domains, as well as the ability to evict those pages from
LLCs, using a capability such as provided by the \clflush instruction
on the x86 architecture.  \clflush is designed to maintain consistency
between caches and memory for write-combined
memory~\cite{guide2010intel}.
The attacker uses \clflush, providing a virtual address as an
argument, to invalidate the cache lines occupied by the backing
physical memory.  After a short time interval (the ``\flushreload
interval'') during which the victim executes, the attacker measures
the time to access the same virtual address.  Based on this duration,
the attacker can infer whether the victim accessed that memory during
the interval.

\subsection{Design}
\label{sec:coa:design}

Modern operating systems, in particular Linux OS, often adopt
on-demand paging and copy-on-write mechanisms~\cite{fabrega1995copy} to
reduce the memory footprints of userspace applications. In particular,
copy-on-write enables multiple processes to share the same set of
physical memory pages as long as none of them modify the content.  If
a process writes to a shared memory page, the write will trigger a
page fault and a subsequent new page allocation so that a private copy
of page will be provided to this process. In addition, memory merging
techniques like \KSM~\cite{arcangeli2009increasing} are also used in
Linux OS to deduplicate identical memory pages.  Memory sharing,
however, is one of the key factors that enable \flushreload side
channel attacks.  Disabling memory page sharing entirely will
eliminate \flushreload side channels but at the cost of much larger
memory footprints and thus inefficient use of physical memory.

\begin{figure}[t]
\includegraphics[width=1.05\linewidth]{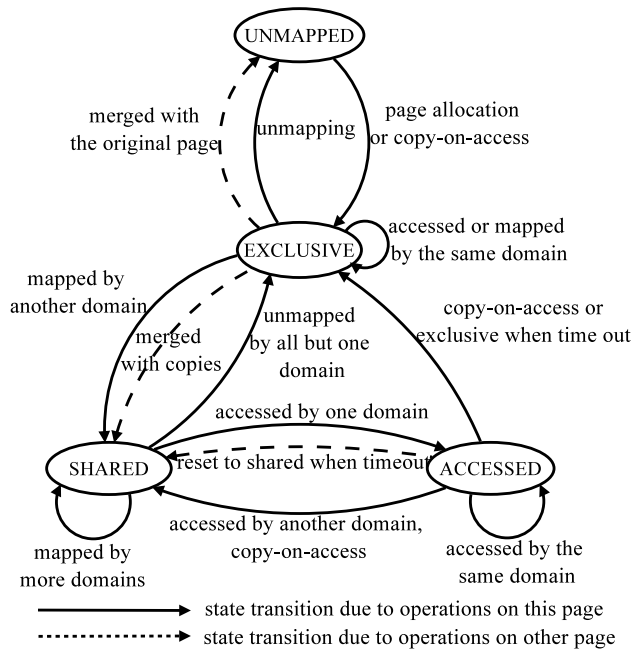}
\caption{State transition of a physical page}
\label{fig:states}
\end{figure}

\sysname adopts a design that we call \coa, which dynamically controls
the sharing of physical memory pages between security domains. We
designate each physical page as being in exactly one of the following
states: \unmapped, \exclusive, \shared, and \accessed.  An \unmapped
page is a physical page that is not currently in use. An \exclusive
page is a physical page that is currently used by exactly one security
domain, but may be shared by one or multiple processes in that domain later.
A \shared page is a physical page that is shared by multiple security
domains, i.e., mapped by at least one process of each of the sharing
domains, but no process in any domain has accessed this physical page
recently. In contrast, an \accessed page is a previously \shared page
that was recently accessed by a security domain.  The state
transitions are shown in \figref{fig:states}.

An \unmapped page can transition to the \exclusive state either due to normal
page mapping, or due to copy-on-access when a page is copied into it.
Unmapping a physical page for any reason (e.g., process
termination, page swapping) will move an \exclusive page back to the \unmapped
state. However, mapping the current \exclusive page by another security domain
will transit it into the \shared state.  If all but one domain unmaps this
page, it will transition back from the \shared state to the \exclusive state, or
\accessed state to the \exclusive state.  A page in the \shared state may be
shared by more domains and remain in the same state; when any one of the domains
accesses the page, it will transition to the \accessed state. An \accessed page
can stay that way as long as only one of security domains accesses it. If this
page is accessed by another domain, a new physical page will be allocated to
make a copy of this one, and the current page will transition to either
\exclusive or \shared state, depending on the remaining number of domains
mapping this page. The new page will be assigned state \exclusive.  An \accessed
page will be reset to the \shared state if it is not accessed for
\accessedTimeout seconds. This timeout mechanism will ensure that only recently
used pages will remain in the \accessed state.  Page merging may also be
triggered by deduplication services in a modern OS (e.g., KSM in Linux).  This
effect is reflected by a dashed line in \figref{fig:states} from state
\exclusive to \shared.  A page at any of the \textit{mapped} states (i.e.,
\exclusive, \shared, \accessed) can transition to \unmapped state for the same
reason when it is a copy of another page (not shown in the figure).

Merging duplicated pages requires some extra bookkeeping. When
a page transitions from \unmapped to \exclusive due to copy-on-access,
the original page is tracked by the new copy so that \sysname knows
with which page to merge it when deduplicating.  If
the original page is unmapped first, then one of its
copies will be designated as the new ``original''
page, with which other copies will be merged in the future. The interaction between copy-on-access
and existing copy-on-write mechanisms is also implicitly depicted in
\figref{fig:states}: Upon copy-on-write, the triggering process will first
\textit{unmap} the physical page, possibly inducing a state transition (from
\shared to \exclusive). The state of the newly mapped physical page is
maintained separately.

\subsection{Implementation}
\label{sec:coa:impl}
At the core of copy-on-access implementation is the state machine depicted in
\figref{fig:states}.

\bheading{\unmapped $\Leftrightarrow$ \exclusive
  $\Leftrightarrow$ \shared}
Conventional Linux kernels maintain the relationship between processes
and the physical pages they use.  However, \sysname also needs to keep
track of the relationship between containers and the physical pages
that the container's processes use.  Therefore, \sysname incorporates
a new data structure, \pcounterincontainer, which is conceptually a
table used for recording, for each physical page, the number of
processes in each container that have Page Table Entries (PTEs) mapped
to this page. Specifically, let
\pcounterincontainers{\containerIdx}{\physPageIdx} indicate the number
of PTEs mapped to physical page \physPageIdx in container
\containerIdx. For example, consider five physical pages and four
containers (see \tabref{tb:ppage_container}).
$\pcounterincontainers{1}{2} = 2$ indicates there are $2$ processes in
container $2$ that have virtual pages mapped to physical page $1$. It
is easy to see from \pcounterincontainer that physical pages $1$, $2$,
and $4$ are in the \shared or \accessed state, physical page $5$ is
\unmapped and physical page $3$ is \exclusive. Of course, the table
\pcounterincontainer needs to be dynamically maintained, as containers
may be created and terminated at any time.  Therefore, to implement
\pcounterincontainer, we added one data field, a pointer to an array
with the size of the number of physical pages in the system, in each
\texttt{PID\_namespace} structure. These data fields collectively
function as the table \pcounterincontainer.

\begin{table}[h]
	\caption{Example of \pcounterincontainer: 5 pages, 4 containers}
	\centering
	{\footnotesize
	\begin{tabular}{|c|c|c|c|c}
		\hline
		\diagbox{physical page}{container ID}& 1&2&3&4\\\hline
		1&0&2&0&1\\\hline
		2&3&1&1&0\\\hline
		3&0&0&1&0\\\hline
		4&1&1&1&1\\\hline
		5&0&0&0&0\\
	\end{tabular}
	}
\label{tb:ppage_container}
\end{table}

The \pcounterincontainer data structure is updated and referenced in
multiple places in the kernel.  Specifically, in \sysname we
instrumented every update of \mapcount, a data field in the
\texttt{page} structure for counting PTE mappings, so that every time
the kernel tracks PTE mapping of a physical page, \pcounterincontainer
is updated accordingly. The use of \pcounterincontainer greatly
simplifies the process of maintaining and determining the state of a
physical page: (1) Given a container, access to a single cell suffices
to check whether a physical page is already mapped in the
container. This operation is very commonly used to decide if a state
transition is required when a page is mapped by a process.  Without
\pcounterincontainer, such a operation requires running through the
entire reverse mapping process and checking whether each mapping is
from the given container. (2) Given a physical page, it takes
\containerNmbr accesses to \pcounterincontainer, where \containerNmbr
is the total number of containers, to determine which containers have
mapped to this page. This operation is commonly used to determine the
state of a physical page. 

\begin{figure}[tb]
\centering
\includegraphics[width=0.95\linewidth]{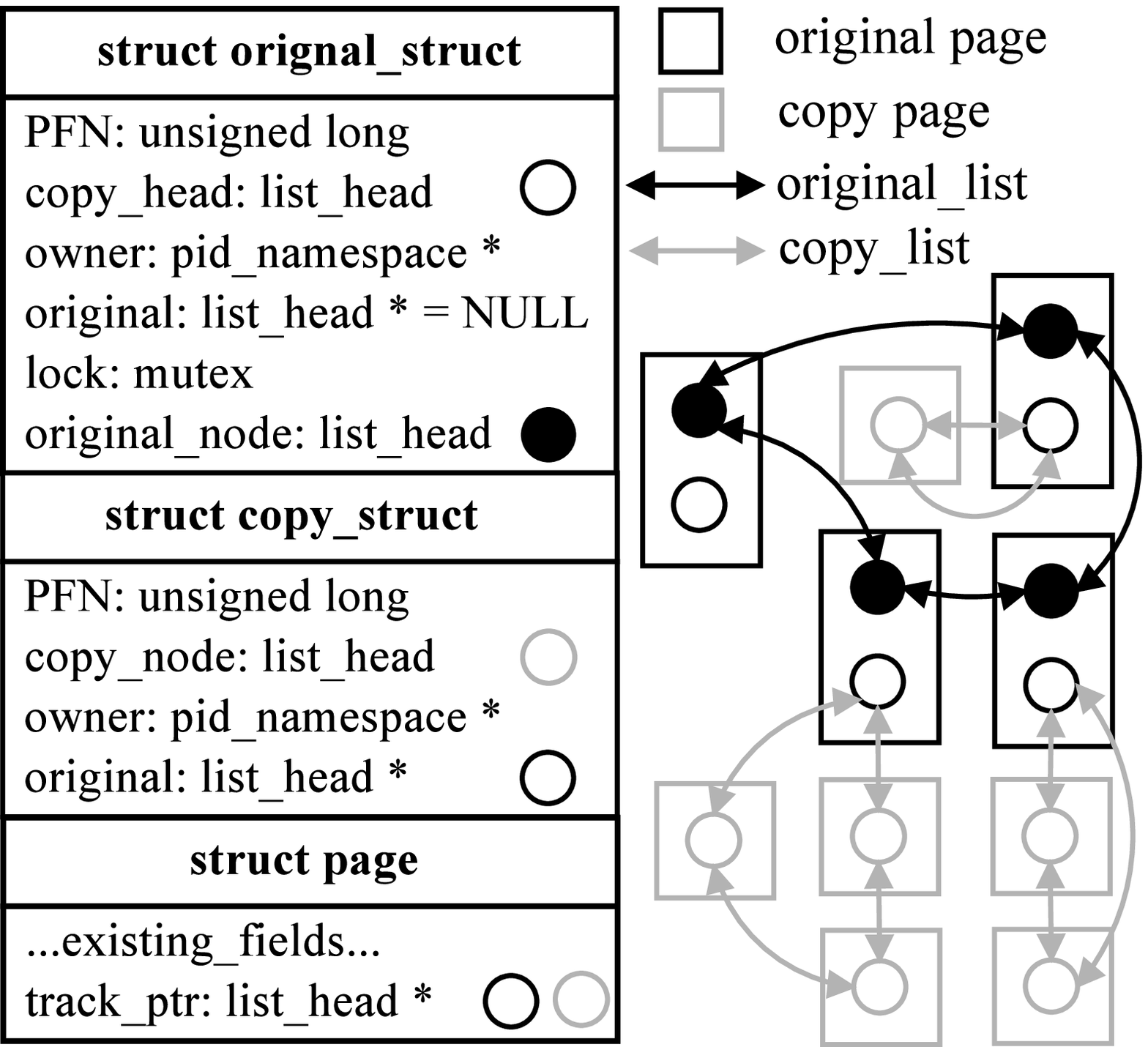}
\caption{Structure of \coa page lists.}
\label{fig:linkedlist}
\end{figure}

\bheading{\shared $\Rightarrow$ \accessed}  
To differentiate \shared and \accessed states, one additional data field,
\owner, is added (see \figref{fig:linkedlist}) to indicate the owner of the page
(a pointer to a \texttt{PID\_namespace} structure). When the page is in the
\shared state, its \owner is \texttt{NULL}; otherwise it points to the
container that last accessed it.

All PTEs pointing to a \shared physical page will have a reserved \COA
bit set. Therefore, any access to these virtual pages will induce a
page fault.  When a page fault is triggered, \sysname checks if the
page is present in physical memory; if so, and if the physical page is
in the \shared state, the \COA bit of the current PTE for this page
will be cleared so that additional accesses to this physical page from
the current process will be allowed without page faults.  The physical
page will also transition to the \accessed state.

\bheading{\accessed $\Rightarrow$ \exclusive/\shared}  If the page is
already in the \accessed state when a domain other than the \owner
accesses it, the page fault handler will allocate a new physical page,
copy the content of the original page into the new page, and change
the PTEs of the processes in the accessing container so that they
point to the new page. Since multiple same-content copies in one
domain burdens both performance and memory but contributes nothing for
security, the fault handler will reuse a copy belonging to that domain
if it exists. After copy-on-access, the original page can either be
\exclusive or \shared.  All copy pages are anonymous-mapped, since
only a single file-mapped page for the same file section is allowed.

A transition from the \accessed state to \shared or \exclusive state
can also be triggered by a timeout mechanism. \sysname implements a
periodic timer (every $\accessedTimeout = 1\secs$). Upon timer
expiration, all physical pages in the \accessedTimeout state that were
not accessed during this \accessedTimeout interval will be reset to
the \shared state by clearing its \owner field, so that pages that are
infrequently accessed are less likely to trigger \coa.  If an
\accessed page is found for which its \pcounterincontainer shows the
number of domains mapped to it is 1, then the daemon instead clears
the \COA bit of all PTEs for that page and marks the page \exclusive.

Instead of keeping a list of \accessed pages, \sysname maintains a list of pages
that are in the \shared or \accessed state, denoted \originalpagelist (shown in
\figref{fig:linkedlist}). Each node in the list also maintains a list of copies
of the page it represents, dubbed \copypagelist, which could be empty.  Both
lists are doubly linked. A tracking pointer \texttt{track\_ptr} to
the \copypagelist or \originalpagelist node is added in the \texttt{struct page} structure to attach
the list onto the array of data structures that represent physical pages.
Whenever a copy is made from the page upon \coa, the copy page is inserted into
the \copypagelist of the original page. Whenever a physical page transitions to
the \unmapped state, it will be removed from whichever of \originalpagelist or
\copypagelist it is contained in.  In the former case, \sysname will designate a
copy page of the original page as the new original page and adjust the lists
accordingly.

Every \accessedTimeout seconds, \sysname will traverse the
\originalpagelist.  If the visited page is in
the \accessed state, the timer interrupt handler
will check whether it has been accessed since the last such check by
seeing if the ACCESSED bit of any PTE for this page in the
\texttt{owner} container is set.  If not, \sysname will transition
this page back to the \shared state.  In any case, the
ACCESSED bit in the PTEs will be cleared.

For security reasons that will be explained in
\secref{sec:coa:security}, we further require flushing the entire
memory page out of the cache after transitioning a page from the
\accessed state to the \shared state due to this timeout
mechanism. This page-flushing procedure is implemented by issuing
\texttt{clflush} on each of the memory blocks of any virtual page that
maps to this physical page.

\bheading{State transition upon \clflush} The \clflush instruction is
subject to the same permission checks as a memory load, will trigger
the same page faults,\footnote{We empirically confirmed this by
  executing \clflush instructions on memory pages with PTE reserved
  bits set.} and will
similarly set the ACCESSED bit in the PTE of its
argument~\cite{guide2010intel}.  As such, each \Flush via \clflush
triggers the same transitions (e.g., from \shared to \accessed, and
from \accessed to an \exclusive copy) as a \Reload in our
implementation, meaning that this defense is equally effective against
both \flushreload and \flushflush~\cite{gruss:2015:FF} attacks.

\bheading{Page deduplication}  To mitigate the impact of \coa on the
size of memory, \sysname implements a less frequent timer (every
$\copyTimeout = 10\times\accessedTimeout$ seconds) to periodically
merge the page copies with their original pages. Within the timer
interrupt handler, \originalpagelist and each \copypagelist are
traversed similarly to the ``\accessed $\Rightarrow$ \shared''
transition description above, though the ACCESSED bit in the PTEs of
only pages that are in the \exclusive state are checked. If a copy
page has not been accessed since the last such check (i.e., the
ACCESSED bit is unset in all PTEs pointing to it), it will be merged
with its original page (the head of the \copypagelist). The ACCESSED
bit in the PTEs will be cleared afterwards.

When merging two pages, if the original page is anonymous-mapped, then
the copy page can be merged by simply updating all PTEs pointing to
the copy page to instead point to the original page, and then updating
the original page's reverse mappings to include these PTEs.  If the
original page is file-mapped, then the merging process is more
intricate, additionally involving the creation of a new virtual memory
area (\texttt{vma} structure) that maps to the original page's file
position and using this structure to replace the virtual memory area
of the (anonymous) copy page in the relevant task structure.

For security reasons, merging of two pages requires flushing the
original physical page from the LLC.  We will elaborate on this point
in \secref{sec:coa:security}.

\bheading{Interacting with KSM}  
Page deduplication can also be triggered by existing memory
deduplication mechanisms (e.g., KSM). To maintain the state of the
physical pages, \sysname instruments every reference to \mapcount
within KSM and updates \pcounterincontainer accordingly. In addition
to merging a copy page with its original, three other types of page
merging in KSM might occur: (1) an original page is merged with a copy
page on a different \copypagelist; (2) a copy page is merged with
another copy page on a different \copypagelist; (3) an original page
is merged with another original page.

\newcommand{\exampleC}{$6$\xspace}
\newcommand{\exampleO}{$1$\xspace}
\newcommand{\exampleCa}{$5$\xspace}
\newcommand{\exampleCb}{$7$\xspace}
\newcommand{\exampleOa}{$3$\xspace}
\newcommand{\exampleOb}{$2$\xspace}
\newcommand{\examplenewcopy}{$9$\xspace}
\newcommand{\exampleneworiginal}{$5$\xspace}
\begin{figure}[t]
\begin{center}
\begin{tabular}{cccc}
	\centering
\subfigure[][Initial state]{
  \label{fig:ksm:initial}
  \includegraphics[width=0.4\linewidth]{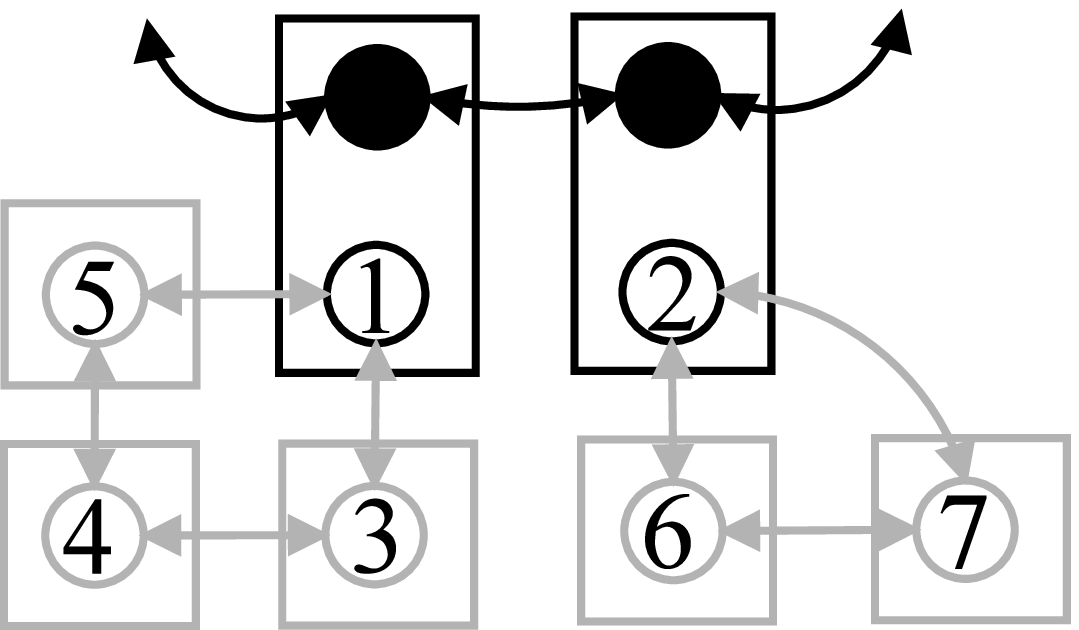}
}
&
\subfigure[][Merging page \exampleCa into \exampleCb]{
  \label{fig:ksm:copyIntoCopy}
  \includegraphics[width=0.4\linewidth]{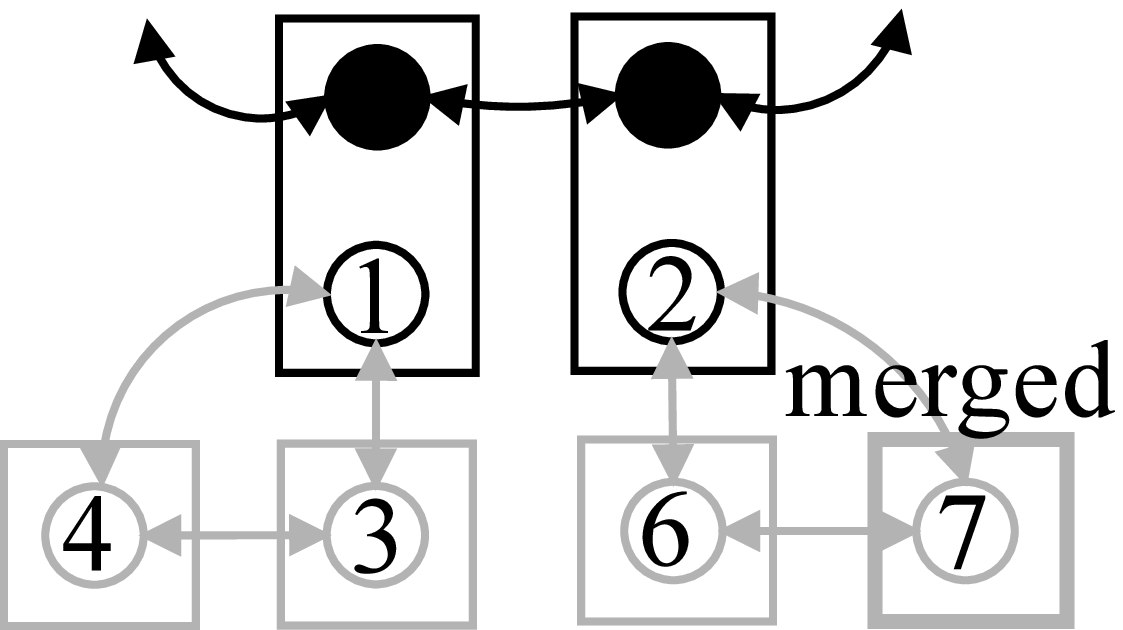}
}
\\
\subfigure[][Merging page \exampleO into \exampleC]{
  \label{fig:ksm:origIntoCopy}
  \includegraphics[width=0.44\linewidth]{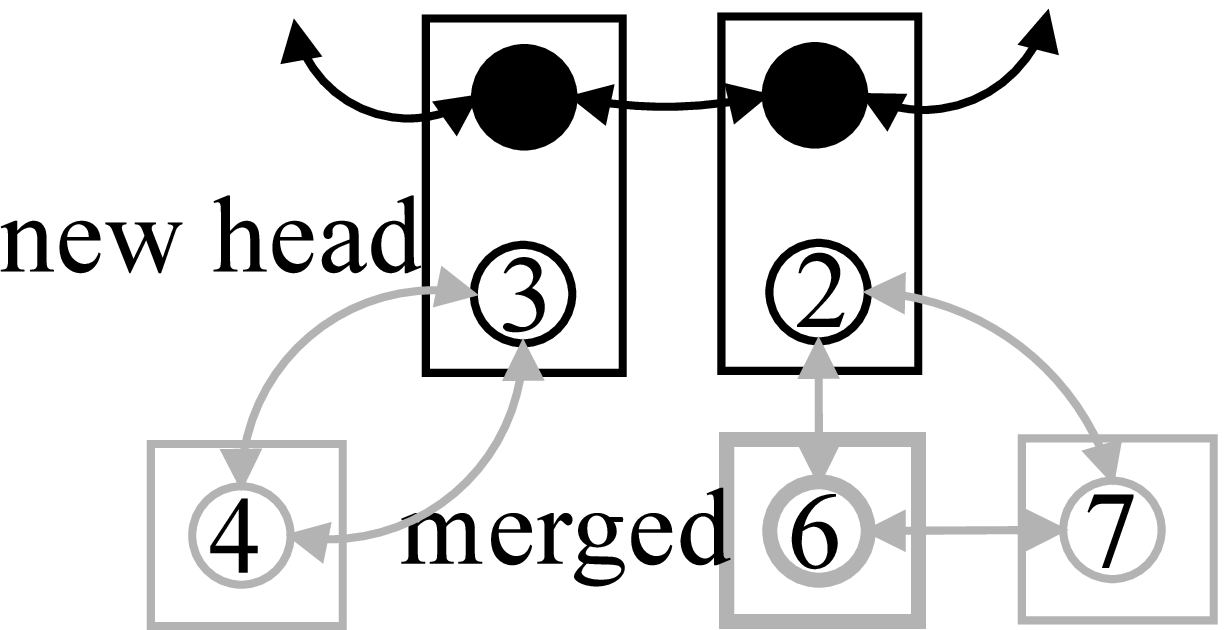}
}
&
\subfigure[][Merging page \exampleOa into \exampleOb]{
  \label{fig:ksm:origIntoOrig}
  \includegraphics[width=0.44\linewidth]{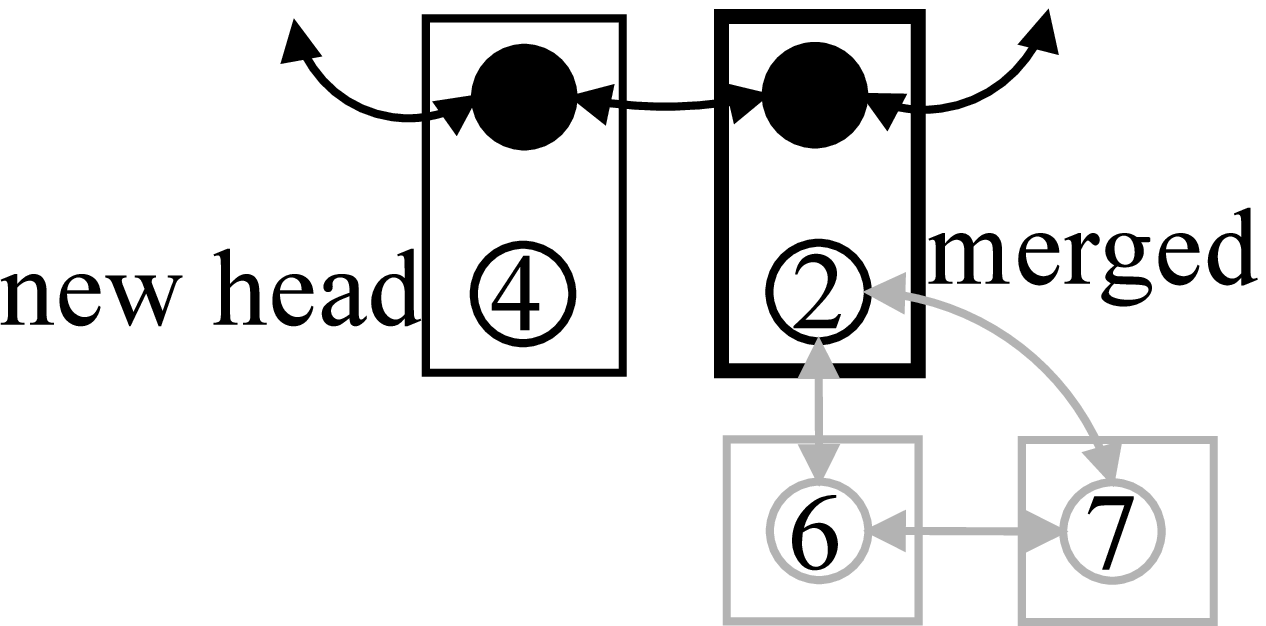}
}
\end{tabular}
\end{center}

\caption{KSM operation example}
\label{fig:ksm}
\end{figure}

We illustrate these operations in the example in \figref{fig:ksm}.
Consider the initial
\originalpagelist and \copypagelist{s} shown in \figref{fig:ksm:initial}, where
all of pages 1--7 have the same content.  \figref{fig:ksm:copyIntoCopy} shows
the list configurations after copy page \exampleCa is merged into copy page
\exampleCb by KSM, after which only page \exampleCb is preserved and is in the
\shared state.  (Copy page \exampleCa is unmapped and removed from
\copypagelist.)  \figref{fig:ksm:origIntoCopy} shows an original page \exampleO
merged into a copy page \exampleC. As a result, page \exampleC becomes a \shared
page, and one of page \exampleO's copies (in our implementation, the first copy
in the list) is designated as the new original page of the list.
\figref{fig:ksm:origIntoOrig} shows original page \exampleOa being merged into
original page \exampleOb.  A copy page in page \exampleOa's \copypagelist
becomes the new original page. A physical page may be disconnected from the
\originalpagelist or any \copypagelist, which we call \textit{untracked}, if it
has never entered \shared or \accessed states. If KSM merges a page that is
untracked with a tracked page, then the untracked page will simply be merged
into the tracked page, which will transition to the \shared state (if not
already there). 

It is apparent that KSM is capable of merging more pages than our built-in page
deduplication mechanisms.  However, \sysname still relies on the built-in page
deduplication mechanisms for several reasons. First, KSM can merge only
anonymous-mapped pages, while \sysname needs to frequently merge an
anonymous-mapped page (a copy) with a file-mapped page (the original). Second,
KSM may not be enabled in certain settings, which will lead to ever growing
\copypagelist{s}.  Third, KSM needs to compare page contents byte-by-byte before
merging two identical pages, whereas \sysname deduplicates pages on the
same \copypagelist, avoiding the expensive page content comparison.

\subsection{Security}
\label{sec:coa:security}

Copy-on-access is intuitively secure by design, as no two security
domains may access the same physical page at the same time, rendering
\flushreload attacks seemingly impossible.  To show security formally,
we subjected our design to \textit{model checking} in order to prove
that \coa is secure against \flushreload attacks.  Model checking is
an approach to formally verify a specification of a finite-state
concurrent system expressed as temporal logic formulas, by traversing
the finite-state machine defined by the model.  In our study, we used
the \spin model checker, which offers efficient ways to model
concurrent systems and verify temporal logic specifications.

Prior works have proposed the use of model checking to evaluate
whether software is vulnerable to remote timing attacks
(e.g.,~\cite{Svenningsson:2009:SVS}).  However, using model checking
to verify that a system is secure against the side-channel attacks of
concern in this paper is, we believe, novel and might be of interest
in its own right.

\bheading{System modeling}
We model a physical page in \figref{fig:states} using a \textit{byte}
variable in the \textsc{Promela} programming language, and two
physical pages as an array of two such variables, named
\texttt{pages}.  We model two security domains (e.g., containers), an
attacker domain and a victim domain, as two processes in
\textsc{Promela}.  Each process maps a virtual page, \mcVirt, to one
of the physical pages. The virtual page is modeled as an index to the
\mcPages{~} array; initially \mcVirt for both the attacker and the
victim point to the first physical page (i.e., \mcVirt is $0$).  The
victim process repeatedly sets \mcPages{\mcVirt} to $1$, simulating a
memory access that brings \mcPages{\mcVirt} into cache.  The attacker
process \Flush{es} the virtual page by assigning $0$ to
\mcPages{\mcVirt} and \Reload{s} it by assigning $1$ to
\mcPages{\mcVirt} after testing if it already equals to $1$.  Both the
\Flush and \Reload operations are modeled as \textit{atomic} to
simplify the state exploration.

We track the state and owner of the first physical page using another
two variables, \mcState and \mcOwner.  The first page is initially in
the \shared state (\mcState is \shared), and state transitions in
\figref{fig:states} are implemented by each process when they access
the memory.  For example, the \Reload code snippet run by the attacker
is shown in \figref{fig:codesnippet}.  If the attacker has access to
the shared page (\lstref{lst:codesnippet:sharedPage}), versus an
exclusive copy (\lstref{lst:codesnippet:exclusivePage}), then it
simulates an access to the page, which either moves the state of the
page to \accessed (\lstref{lst:codesnippet:toAccessed}) if the state
was \shared (\lstref{lst:codesnippet:testShared}) or to \exclusive
(\lstref{lst:codesnippet:toExclusive}) after making a copy
(\lstref{lst:codesnippet:coa}) if the state was already \accessed and
not owned by the attacker (\lstref{lst:codesnippet:testAccessed}).
Leakage is detected if \mcPages{\mcVirt} is $1$ prior to the
attacker setting it as such (\lstref{lst:codesnippet:retrieve}), which
the attacker tests in \lstref{lst:codesnippet:leakage}.

\lstset{language=Python, frame=lines, numbers=left, stepnumber=1,
  basicstyle=\scriptsize\ttfamily, keywordstyle=\scriptsize\ttfamily,
  numberstyle=\scriptsize, tabsize=2, breaklines=true, showstringspaces=false,
  xleftmargin=18pt, escapeinside={(*}{*)}, numberblanklines=false}

\begin{figure}[h]
\begin{lstlisting}
atomic {
   if
   ::(virt==0) ->  (*\label{lst:codesnippet:sharedPage}*)
      if
      ::(state==UNMAPPED) -> (*\label{lst:codesnippet:testUnmapped}*)
         assert(0)
      ::(state==EXCLUSIVE && owner!=ATTACKER) ->(*\label{lst:codesnippet:testExclusive}*)
         assert(0)
      ::(state==SHARED) -> (*\label{lst:codesnippet:testShared}*)
         state=ACCESSED  (*\label{lst:codesnippet:toAccessed}*)
         owner=ATTACKER  (*\label{lst:codesnippet:chngOwner}*)
      ::(state==ACCESSED && owner!=ATTACKER) -> (*\label{lst:codesnippet:testAccessed}*)
         virt=1 /* copy-on-access */ (*\label{lst:codesnippet:coa}*)
         state=EXCLUSIVE  (*\label{lst:codesnippet:toExclusive}*)
      fi
   ::else -> skip (*\label{lst:codesnippet:exclusivePage}*)
   fi
   assert(pages[virt]==0) (*\label{lst:codesnippet:leakage}*)
   pages[virt]=1 (*\label{lst:codesnippet:retrieve}*)
}
\end{lstlisting}
\caption{Code snippet for \Reload. The procedures for other memory
accesses are similar.}
\label{fig:codesnippet}
\end{figure}

To model the dashed lines in \figref{fig:states}, we implemented
another process, called \textit{timer}, in \textsc{Promela} that
periodically transitions the physical page back to \shared state from
\accessed state, and periodically with a longer interval, merges the
two pages by changing the value of \mcVirt of each domain back to $0$,
\mcOwner to \mcNoOwner, and \mcState to \shared.

The security specification is stated as a non-interference property.
Specifically, as the attacker domain always first \Flush{es} the memory
block (sets \mcPages{\mcVirt} to $0$) before \Reload{ing} it (setting
\mcPages{\mcVirt} to $1$), if the non-interference property holds,
then it should follow that the attacker should always
find \mcPages{\mcVirt} to be $0$ upon \Reload{ing} the page.  The model
checker checks for violation of this property in the verification.

\bheading{Automated verification}  We checked the model using
\spin. Interestingly, our first model-checking attempt suggested that
the state transitions may leak information to a \flushreload
attacker. The leaks were caused by the \textit{timer} process that
periodically transitions the model to a \shared state. After
inspecting the design and implementation, we found that there were two
situations that may cause information leaks.  In the first case, when
the timer transitions the state machine to the \shared state from the
\accessed state, if the prior owner of the page was the victim and the
attacker reloaded the memory right after the transition, the attacker
may learn one bit of information. In the second case, when the
physical page was merged with its copy, if the owner of the page was
the victim before the page became \shared, the attacker may reload it
and again learn one bit of information.  Since in our implementation
of \sysname, these two state transitions are triggered if the page (or
its copy) has not been accessed for a while (roughly \accessedTimeout
and \copyTimeout seconds, respectively), the information leakage
bandwidth due to each would be approximately $1/\accessedTimeout$ bits
per page per second or $1/\copyTimeout$ bits per page per second,
respectively.

We improved our \sysname implementation to prevent this leakage by enforcing LLC
flushes (as described in \secref{sec:coa:impl}) upon these two periodic state
transitions.  We adapted our model accordingly to reflect such changes by adding
one more instruction to assign \mcPages{0} to be $0$ right after the two
\textit{timer}-induced state transitions.  Model checking this refined model
revealed no further information leakage in the design.

\section{Cacheability Management for \primeprobe Defense}
\label{sec:cacheabilityMgmt}

Another common method to launch side-channel attacks via caches is using
\primeprobe attacks, introduced by Osvik et al.~\cite{osvik2006cache}.  These
attacks have recently been adapted to use LLCs to great effect,
e.g.,~\cite{liu2015practical, irazoqui2015shared}.  Unlike a \flushreload
attack, \primeprobe attacks do not require the attacker and victim security
domains to share pages.  Rather, the attacker simply needs to access memory so
as to evict (\Prime) the contents of a cache set and later access (\Probe) this
memory again to determine (by timing the accesses) how much the victim evicted
from the cache set.  A potentially effective countermeasure to these attacks,
accordingly, is to remove the attacker's ability to \Prime and \Probe the whole
cache set and to predict how a victim's demand for that set will be reflected in
the number of evictions from that set.

\subsection{Design}
\label{sec:cacheabilityMgmt:design}

Suppose a \cacheLineNmbr-way set associative LLC, i.e., so that each
cache set has \cacheLineNmbr lines.  Let \evictedNmbr be the number
of cache lines in one set that the attacker observes having been
evicted in a \primeprobe interval.  The \primeprobe attack is
effective today because \evictedNmbr is typically a good indicator of
the demand \victimDemand that the victim security domain had for
memory that mapped to that cache set during the \primeprobe interval.
In particular, if the attacker \Prime{s} and \Probe{s} all
\cacheLineNmbr lines, then it can often observe the victim's demand
\victimDemand exactly, unless $\victimDemand > \cacheLineNmbr$ (in
which case the attacker learns at least $\victimDemand \ge \cacheLineNmbr$).

The alternative that we propose here is to periodically and
probabilistically reconfigure the budget
\linesPerContainer{\containerIdx} of lines per cache set that the
security domain \containerIdx can occupy.  After such a
reconfiguration, the attacker's view of the victim's demand
\victimDemand is clouded by the following three effects.  First, if
the attacker is allotted a budget $\linesPerContainer{\attackerLabel}
< \cacheLineNmbr$, then the attacker will be unable to observe any
evictions at all (i.e., $\evictedNmbr=0$) if $\victimDemand <
\cacheLineNmbr-\linesPerContainer{\attackerLabel}$.\footnote{This
  statement assumes a least-recently-used replacement policy and that
  the victim is the only security domain that runs in the \primeprobe
  interval.  If it was not the only security domain to run and
  evictions were caused by another, then the ambiguity of what caused
  the observable evictions will additionally cause difficulties for
  the attacker.}  Second, if the victim is given allotment
\linesPerContainer{\victimLabel}, then any two victim demands
\victimDemand, \victimDemandAlt satisfying $\victimDemand >
\victimDemandAlt \ge \linesPerContainer{\victimLabel}$ will be
indistinguishable to the attacker.  Third, the probabilistic
assignment of \linesPerContainer{\victimLabel} results in extra
ambiguity for the attacker, since \evictedNmbr evictions might reflect
the demand \victimDemand or the budget
\linesPerContainer{\victimLabel}, since $\evictedNmbr \le
\min\{\victimDemand,\linesPerContainer{\victimLabel}\}$ (if all
\evictedNmbr evictions are caused by the victim).

To enforce the budget \linesPerContainer{\containerIdx} of lines that
security domain \containerIdx can use in a given cache set, \sysname
maintains for each cache set a queue per security domain that records
which memory blocks are presently cacheable in this set by processes
in this domain.  Each element in the queue indicates a memory block
that maps to this cache set; only blocks listed in the queue can be
cached in that set.  The queue is maintained with a least recently
used (LRU) replacement algorithm. That is, whenever a new memory block
is accessed, it will replace the memory block in the corresponding
queue that is the least recently used.

\begin{figure*}[t]
  \begin{minipage}[c]{0.70\textwidth}
    \includegraphics[width=0.95\textwidth]{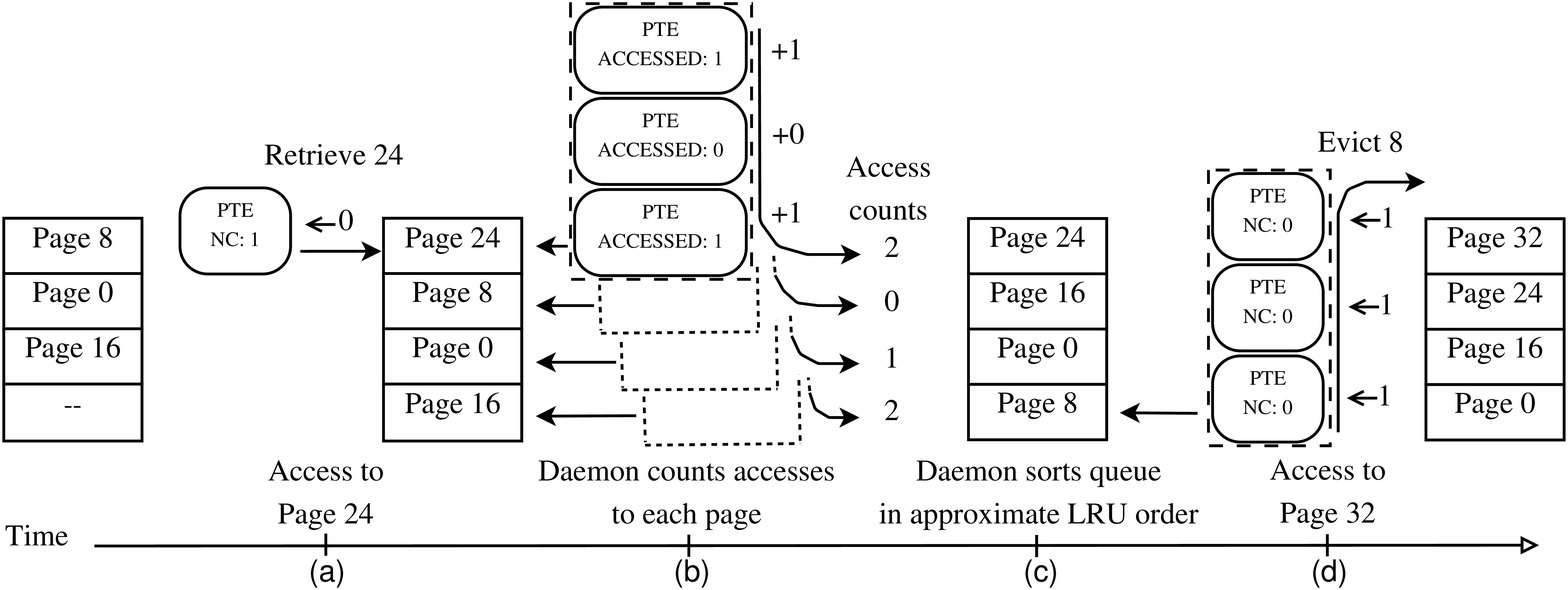}
  \end{minipage}\hfill
  \begin{minipage}[c]{0.30\textwidth}
    \caption{A \lru for one page color in a domain: (a) access to page 24
    brings the page into the queue and clears the NC bit
    (``$\leftarrow 0$'') in the PTE triggering the fault; periodically,
    (b) a daemon counts, per page, the ACCESSED bits (``$+0$'',
    ``$+1$'') in the domain's PTEs referring to that page and (c)
    reorders the pages in the queue accordingly; to make room for a
    new page, (d) the NC bits in PTEs pointing to the least recently
    used page are set (``$\leftarrow 1$''), and the page is removed
    from the queue.}
    \label{fig:queues:all}
  \end{minipage}
\end{figure*}

\subsection{Implementation}
\label{sec:cacheabilityMgmt:impl}

Implementation of \lru{s} is processor micro-architecture
dependent.  Here we focus our attention on Intel x86 processors, which
appears to be more vulnerable to \primeprobe attacks due to their
inclusive last-level cache~\cite{liu2015practical}.  As x86
architectures only support memory management at the page granularity
(e.g., by manipulating the PTEs to cause page faults), \sysname
controls the cacheability of memory blocks at page granularity.
\sysname uses reserved bits in each PTE to manage the cacheability of,
and to track accesses to, the physical page to which it points, since
a reserved bit set in a PTE induces a page fault upon access to the
associated virtual page, for which the backing physical page cannot be
retrieved or cached (if it is not already) before the bit is
cleared~\cite{guide2010intel,raikin2014tracking}.
We hence use the term \textit{domain-cacheable} to refer to a physical
page that is ``cacheable'' in the view of all processes in a
particular security domain, which is implemented by modifying all
relevant PTEs (to have no reserved bits set) in the processes of that
security domain. By definition, a physical page that is
domain-cacheable to one container may not necessarily be
domain-cacheable to another.

To ensure that no more than \linesPerContainer{\containerIdx} memory
blocks from all processes in container \containerIdx can occupy lines
in a given cache set, \sysname ensures that no more than
\linesPerContainer{\containerIdx} of those processes' physical memory
pages, of which contents can be stored in that cache set, are
domain-cacheable at any point in time. Physical memory pages of which
contents can be stored in the same cache set are said to be of the
same \textit{color}, and so to implement this property, \sysname
maintains, per container and per color (rather than per cache set),
one \lru, each element of which is a physical memory page that is
domain-cacheable in this container.
Since the memory blocks in each physical page map to different cache
sets,\footnote{In this way, the cache contention arising from accesses
  in the same page are minimized.} limiting the domain-cacheable pages
of a color to \linesPerContainer{\containerIdx} also limits the number
of cache lines that blocks from these pages can occupy in the same
cache set to \linesPerContainer{\containerIdx}.

To implement a non-domain-cacheable memory page, \sysname sets one of
the reserved bits, which we denote by \LRU, in the PTE for each
virtual page in the domain mapped to that physical page.  As such,
accesses to any of these virtual pages will be trapped into the kernel
and handled by the page fault handler. Upon detecting page faults of
this type, the page fault handler will move the accessed physical page
into the corresponding \lru, clear the \LRU bit in the current
PTE\footnote{We avoid the overhead of traversing all PTEs in the
  container that map to this physical page. Access to those virtual
  pages will trigger page faults to make these updates without
  altering the \lru.}, and remove a least recently used physical page
from the \lru and set the \LRU bits in this domain's PTEs mapped to
that page.  A physical page removed from the \lru will be flushed out
of the cache using \clflush instructions on all of its memory blocks
to ensure that no residue remains in the cache. \sysname will flush
the translation lookaside buffers (TLB) of all processors to ensure
the correctness of page cacheabilities every time PTEs are altered. In
this way, \sysname limits the number of domain-cacheable pages of a
single color at any time to \linesPerContainer{\containerIdx}.

To maintain the LRU property of the \lru, a daemon periodically
re-sorts the queue in descending order of recent access count.
Specifically, the daemon traverses the domain's
PTEs mapped to the \ppage within that domain's queue and counts the number having their ACCESSED bit set, after which
it clears these ACCESSED bits.  It then orders the physical pages in
the \lru by this count (see \figref{fig:queues:all}).  In our present
implementation, this daemon is the same daemon that resets pages from
the \accessed state to \shared state (see \secref{sec:coa}), which
already checks and resets the ACCESSED bits in copies' PTEs.  Again, this
daemon runs every $\accessedTimeout = 1\secs$ seconds in our
implementation.  This daemon also performs the task of resetting
\linesPerContainer{\containerIdx} for each security domain
\containerIdx, which in our present implementation it does every tenth
time it runs.

\begin{figure}[bt]
	\centering
\includegraphics[width=\linewidth]{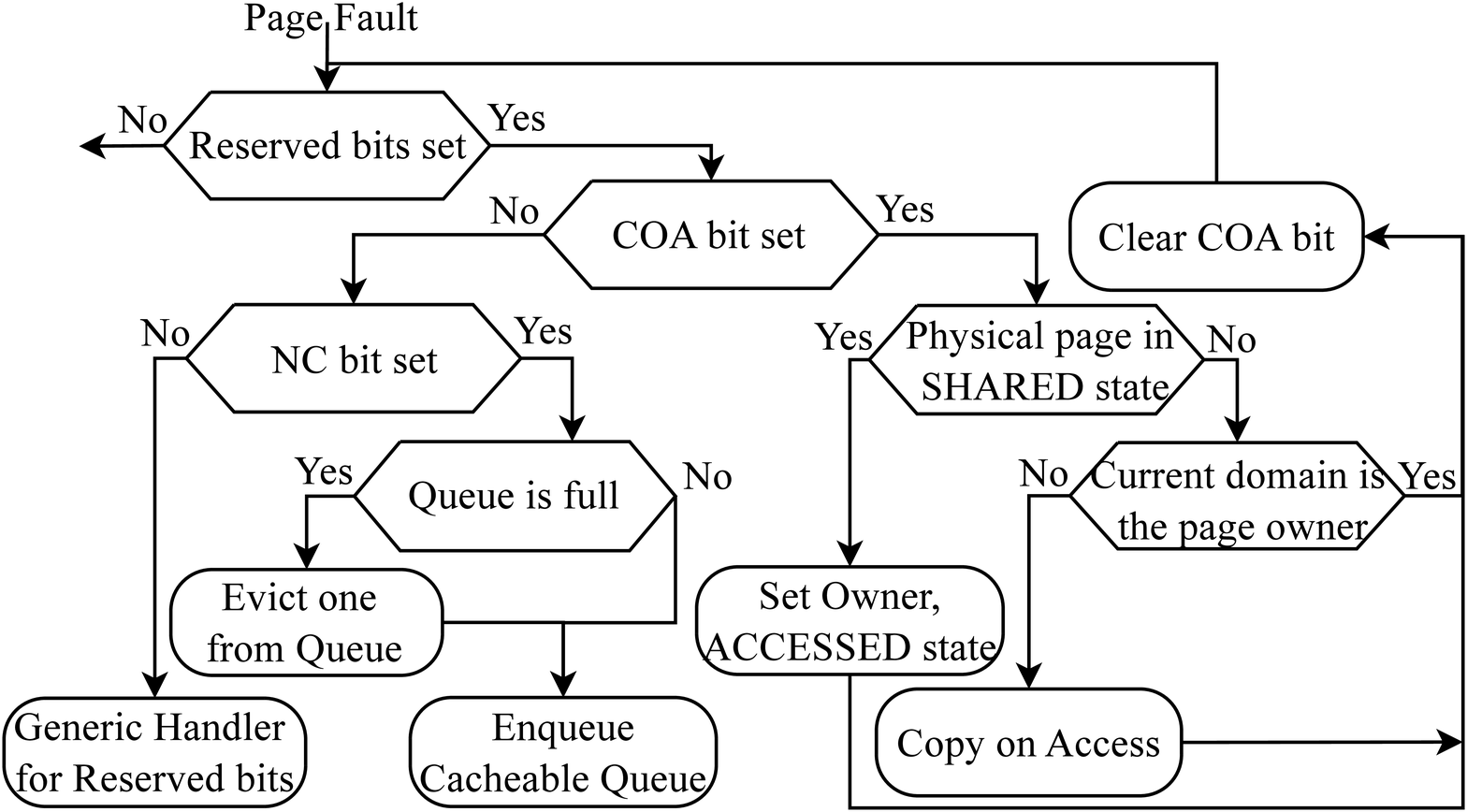}
\label{fig:fault}
\caption{Page fault handler for \sysname.}
\end{figure}

\bheading{Interacting with copy-on-access} The \lru{s} work closely
with the copy-on-access mechanisms. In particular, as both the \COA
and \LRU bits may trigger a page fault upon page accesses, the page
handler logic must incorporate both (shown in Fig. 6). First, a page
fault is handled as normal unless it is due to one of the reserved
bits set in the PTE. As \sysname is the only source of reserved bits,
it takes over page fault handling from this point. \sysname first
checks the \COA bit in the PTE. If it is set, the corresponding
physical page is either \shared, in which case it will be transitioned
to \accessed, or \accessed, in which case it will be copied.  \sysname
then clears the \COA bit and, if no other reserved bits are set, the
fault handler returns.  Otherwise, if the \LRU bit is set, the
associated physical page is not in the \lru for its domain, and so
\sysname enqueues the page and, if the queue is full, removes the
least-recently-used page from the queue.  If the \LRU bit is clear,
this page fault is caused by unknown reasons and \sysname turns
control over to the generic handler for reserved bits.

\subsection{Security}
\label{sec:cacheabilityMgmt:security}

Recall that \linesPerContainer{\containerIdx} is the number of cache
lines in a certain cache set that is made available to security domain
\containerIdx for a period of time.  While the budget
\linesPerContainer{\containerIdx} is in effect, each access to a
memory block that maps to this cache set, beyond the in-queue
\linesPerContainer{\containerIdx} memory blocks, will incur a page
fault (because they are all in different pages).  Because the
page-fault processing time will overwhelm the timing granularity of
modern \primeprobe attacks by an order of magnitude, the attacker
\containerIdx realistically needs to restrict himself to accessing
\linesPerContainer{\containerIdx} pages in his \Probe phase and hence
to occupying \linesPerContainer{\containerIdx} lines in that cache
set.

The security of this design hinges critically on how each
\linesPerContainer{\containerIdx} is set by the daemon.  When
\linesPerContainer{\containerIdx} is reset, it is drawn from a
distribution.  In the remainder of this section we present how this
distribution is determined.

Suppose there are (at most) \attackerNmbr security domains on a host
that are owned by the attacker---which might be all security domains
on the host except the victim---and let \cacheLineNmbr be the number of
LLC cache lines per cache set.  Below we consider security domain $0$
to be the ``victim'' domain being subjected to \primeprobe attacks by
the ``attacker'' domains $1, \ldots, \attackerNmbr$.  Of course, the
attacker domains make use of all
$\sum_{\containerIdx=1}^{\attackerNmbr}
\linesPerContainer{\containerIdx}$ cache lines available to them for
conducting their \primeprobe attacks.

Periodically, \sysname draws a new value
\linesPerContainer{\containerIdx} for each security domain
\containerIdx.  This drawing is memoryless and independent of the
draws for other security domains.  Let
\linesPerContainerRV{\containerIdx} denote the random variable
distributed according to how \linesPerContainer{\containerIdx} is
determined.  The random variables that we presume can be observed by
the attacker domains include $\linesPerContainerRV{1}, \ldots,
\linesPerContainerRV{\attackerNmbr}$; let $\attackerLineNmbrRV{} =
\min\left\{\cacheLineNmbr, \sum_{\containerIdx=1}^{\attackerNmbr}
\linesPerContainerRV{\containerIdx}\right\}$ denote the number of
cache lines allocated to the attacker domains.  We also presume that
the attacker can accurately measure the number \evictedNmbrRV of the
attacker's cache lines that are evicted during the victim's execution.

Let \prob{\victimDemand}{\genericEvent} denote the probability of
event \genericEvent in an execution period during which the victim's
cache usage would populate \victimDemand lines (of this color) if it
were allowed to use all \cacheLineNmbr lines, i.e., if
$\linesPerContainer{0} = \cacheLineNmbr$.  We (the defender) would
like to distribute $\linesPerContainerRV{0}, \ldots,
\linesPerContainerRV{\attackerNmbr}$ and thus \attackerLineNmbrRV so
as to minimize the statistical distance between eviction distributions
observable by the attacker for different victim demands
\victimDemand, \victimDemandAlt, i.e., to minimize
\begin{align}
\sum_{0 \le \victimDemand < \victimDemandAlt \le \cacheLineNmbr} \sum_{\evictedNmbr} |\prob{\victimDemand}{\evictedNmbrRV = \evictedNmbr} - \prob{\victimDemandAlt}{\evictedNmbrRV = \evictedNmbr}|
\label{eqn:securityGoal}
\end{align}

We begin by deriving an expression
for \prob{\victimDemand}{\evictedNmbrRV = \evictedNmbr}.  Below we
make the conservative assumption that all evictions are caused by the
victim's behavior; in reality, caches are far noisier.  We first
consider the case $\evictedNmbr=0$, i.e., that the attacker domains
observe no evictions.
\begin{align*}
\cprob{\Big}{\victimDemand}{\evictedNmbrRV = 0}{\hspace{-0.5em}\begin{array}{l}\linesPerContainerRV{0} = \linesPerContainer{0} \\ \wedge~\attackerLineNmbrRV = \attackerLineNmbr\end{array}\hspace{-0.5em}}
& \!=\! \left\{\begin{array}{@{\extracolsep{-0.4em}}ll}
1 & \mbox{if $\cacheLineNmbr \ge \attackerLineNmbr + \min\{\linesPerContainer{0}, \victimDemand\}$} \\
0 & \mbox{otherwise}
\end{array}\right.
\end{align*}
``$\min\{\linesPerContainer{0}, \victimDemand\}$'' is used above
because any victim demand for memory blocks that map to this cache set
beyond \linesPerContainer{0} will back-fill the cache lines invalidated
when \sysname flushes other blocks from the victim's cacheability
queue, rather than evicting others.  Since \linesPerContainerRV{0} and
\attackerLineNmbrRV are distributed independently,
\begin{align}
\prob{\victimDemand}{\evictedNmbrRV = 0}
& = \sum_{\linesPerContainer{0}=0}^{\victimDemand} \sum_{\attackerLineNmbr=0}^{\cacheLineNmbr-\linesPerContainer{0}} \prob{}{\linesPerContainerRV{0} = \linesPerContainer{0}} \cdot \prob{}{\attackerLineNmbrRV = \attackerLineNmbr} \nonumber \\
& \hspace{1em} + \hspace{-0.75em} \sum_{\linesPerContainer{0}=\victimDemand+1}^{\cacheLineNmbr} \sum_{\attackerLineNmbr=0}^{\cacheLineNmbr-\victimDemand} \prob{}{\linesPerContainerRV{0} = \linesPerContainer{0}} \cdot \prob{}{\attackerLineNmbrRV = \attackerLineNmbr} \label{eqn:zeroEvictions}
\end{align}
Note that we have dropped the ``\victimDemand'' subscripts from the
probabilities on the right, since \linesPerContainerRV{0}
and \attackerLineNmbrRV are distributed independently
of \victimDemand.  And, since
$\linesPerContainerRV{1},\ldots,\linesPerContainerRV{\attackerNmbr}$
are independent, for $\attackerLineNmbr < \cacheLineNmbr$,
\begin{align}
\prob{}{\attackerLineNmbrRV = \attackerLineNmbr} =
\sum_{\substack{\linesPerContainer{1},\ldots,\linesPerContainer{\attackerNmbr}:\\ \linesPerContainer{1} + \ldots + \linesPerContainer{\attackerNmbr} = \attackerLineNmbr}} \prod_{\containerIdx=1}^{\attackerNmbr} \prob{}{\linesPerContainerRV{\containerIdx} = \linesPerContainer{\containerIdx}}
\label{eqn:attackerLineNmbr}
\end{align}
and
\begin{align}
\prob{}{\attackerLineNmbrRV = \cacheLineNmbr} =
\sum_{\substack{\linesPerContainer{1},\ldots,\linesPerContainer{\attackerNmbr}:\\ \linesPerContainer{1} + \ldots + \linesPerContainer{\attackerNmbr} \ge \cacheLineNmbr}} \prod_{\containerIdx=1}^{\attackerNmbr} \prob{}{\linesPerContainerRV{\containerIdx} = \linesPerContainer{\containerIdx}}
\label{eqn:attackerLineNmbrAll}
\end{align}
Similarly, for $\evictedNmbr \ge 1$, 
\begin{align*}
\cprob{\Big}{\victimDemand}{\evictedNmbrRV = \evictedNmbr}{\hspace{-0.5em}\begin{array}{l}\linesPerContainerRV{0} = \linesPerContainer{0} \\ \wedge~\attackerLineNmbrRV = \attackerLineNmbr\end{array}\hspace{-0.5em}}
& = \left\{\begin{array}{@{\extracolsep{-0.4em}}ll}
1 & \mbox{if $\evictedNmbr\! +\!\cacheLineNmbr~=$} \\
  & \mbox{\hspace{1em} $\attackerLineNmbr\! + \!\min\{\linesPerContainer{0}, \victimDemand\}$} \\
0 & \mbox{otherwise}
\end{array}\right.
\end{align*}
and so for $\evictedNmbr \ge 1$,
\begin{align}
\prob{\victimDemand}{\evictedNmbrRV = \evictedNmbr}
& = \sum_{\linesPerContainer{0}=0}^{\victimDemand} \prob{}{\linesPerContainerRV{0} = \linesPerContainer{0}} \cdot \prob{}{\attackerLineNmbrRV = \evictedNmbr\! +\! \cacheLineNmbr \!-\! \linesPerContainer{0}} \nonumber \\
& \hspace{1em} + \hspace{-0.75em} \sum_{\linesPerContainer{0}=\victimDemand+1}^{\cacheLineNmbr} \hspace{-0.53em}\prob{}{\linesPerContainerRV{0} = \linesPerContainer{0}} \cdot \prob{}{\attackerLineNmbrRV = \evictedNmbr\! +\! \cacheLineNmbr \!-\! \victimDemand}
\label{eqn:moreThanZeroEvictions}
\end{align}

From here, we proceed to solve for the best distribution for
$\linesPerContainerRV{0}, \ldots, \linesPerContainerRV{\attackerNmbr}$
to minimize \eqnref{eqn:securityGoal} subject to
constraints \eqnsref{eqn:zeroEvictions}{eqn:moreThanZeroEvictions}.
That is, we specify
constraints \eqnsref{eqn:zeroEvictions}{eqn:moreThanZeroEvictions},
along with
\begin{align}
\forall \containerIdx, \containerIdxAlt, \linesPerContainer{}:~ &
\prob{}{\linesPerContainerRV{\containerIdx} = \linesPerContainer{}} = \prob{}{\linesPerContainerRV{\containerIdxAlt} = \linesPerContainer{}} \label{eqn:allDistsSame}\\
\forall \containerIdx:~ &
\sum_{\linesPerContainer{\containerIdx} = 0}^{\cacheLineNmbr} \prob{}{\linesPerContainerRV{\containerIdx} = \linesPerContainer{\containerIdx}} = 1 \\
\forall \containerIdx, \linesPerContainer{\containerIdx}:~ &
\prob{}{\linesPerContainerRV{\containerIdx} = \linesPerContainer{\containerIdx}} \ge 0 \label{eqn:nonnegProbs}
\end{align}
and then solve for each \prob{}{\linesPerContainerRV{\containerIdx}
= \linesPerContainer{\containerIdx}} to
minimize \eqnref{eqn:securityGoal}.

Unfortunately, solving to minimize \eqnref{eqn:securityGoal} alone
simply results in a distribution that results in no use of the cache
at all (e.g., $\prob{}{\linesPerContainerRV{\containerIdx}=0} = 1$ for
each \containerIdx).  As such, we need to rule out such degenerate
and ``unfair'' cases:
\begin{align}
\forall \containerIdx:~ &
\prob{}{\linesPerContainerRV{\containerIdx}
        < \cacheLineNmbr/(\attackerNmbr+1)} = 0
\label{eqn:fairness}
\end{align}
Also, to encourage cache usage, we counterbalance
\eqnref{eqn:securityGoal} with a second optimization criterion that
values greater use of the cache.  We express this goal as minimizing
the earth mover's distance~\cite{mallows1972emd,elizaveta2001emd} from
the distribution that assigns
$\prob{}{\linesPerContainerRV{\containerIdx} = \cacheLineNmbr} = 1$,
i.e.,
\begin{align}
\sum_{\linesPerContainer{}=0}^{\cacheLineNmbr} (\cacheLineNmbr-\linesPerContainer{})
\cdot \prob{}{\linesPerContainerRV{0} = \linesPerContainer{}}
\label{eqn:performanceGoal}
\end{align}

As such, our final optimization problem seeks to balance
\eqnref{eqn:securityGoal} and \eqnref{eqn:performanceGoal}.  Let
constant \maxSecValue denote the maximum (i.e., worst) possible value
of \eqnref{eqn:securityGoal} (i.e., when
$\prob{}{\linesPerContainerRV{\containerIdx}=\cacheLineNmbr} = 1$ for
each \containerIdx) and \maxPerfValue denote the maximum (i.e., worst)
possible value of \eqnref{eqn:performanceGoal} (i.e., when
$\prob{}{\linesPerContainerRV{\containerIdx}=0} = 1$ for
each \containerIdx).  Then, given a parameter \balanceSlack, $0
< \balanceSlack < 1$, our optimization computes
distributions
for $\linesPerContainerRV{0}, \ldots, \linesPerContainerRV{\attackerNmbr}$
so as to minimize a value \balanceValue subject to
\begin{align*}
\balanceValue & = \frac{1}{\maxSecValue} \left(\sum_{0 \le \victimDemand < \victimDemandAlt \le \cacheLineNmbr} \sum_{\evictedNmbr} |\prob{\victimDemand}{\evictedNmbrRV = \evictedNmbr} - \prob{\victimDemandAlt}{\evictedNmbrRV = \evictedNmbr}|\right) \\
\balanceValue & \ge \frac{1}{\maxPerfValue(1+\balanceSlack)} \left(\sum_{\linesPerContainer{}=0}^{\cacheLineNmbr} (\cacheLineNmbr-\linesPerContainer{})
\cdot \prob{}{\linesPerContainerRV{0} = \linesPerContainer{}}\right)
\end{align*}
and constraints \eqnsref{eqn:zeroEvictions}{eqn:fairness}.

Our evaluation in \secref{sec:eval:security:primeprobe} and
\secref{sec:eval:perf} empirically characterizes the security and
performance that result from setting $\balanceSlack = 0.01$ the
default setting in \sysname.  Of course, other balances could be
chosen between these concerns, though as we will see below, this
setting achieves convincing security while inducing only a modest
performance overhead for most PaaS workloads.

\section{Evaluation}
\label{sec:eval}

In this section, we evaluate the security and performance of \sysname
to validate its design and implementation.

\subsection{Setup}
\label{sec:eval:setup}

Our testbed is a rack mounted DELL server equipped with two 2.67\ghertz Intel Xeon
5550 processors. Each processor contains 4 physical cores (hyperthreading
disabled) sharing an 8\mbytes last-level cache (L3). Each core has a
32\kbytes L1 data and instruction cache and a 256\kbytes L2 unified cache.  The
rack server is equipped with 128\gbytes DRAM and 1000\mbps NIC connected to a
1000\mbps ethernet.

We implemented \sysname as a kernel extension based on the Linux
kernel (version 3.13.11.6) that runs a Ubuntu 14.04 server
edition. Our implementation adds around 7000 lines of code to this
mainstream Linux kernel. We set up containers using \docker (version
1.7.1).

\subsection{Security Evaluation}
\label{sec:eval:security}

We empirically evaluated the effectiveness of \sysname in defending
against both \flushreload and \primeprobe attacks.

\subsubsection{\flushreload Attacks}
\label{sec:eval:security:flushreload}

Although we used \spin model checker to validate the security
of our copy-on-access design (\secref{sec:coa}), we empirically tested
our implementation to validate its effectiveness.  To do so, we
constructed a \flushreload{-based} covert channel between sender and
receiver processes, which were isolated in different containers. Both
the sender and receiver were linked to a shared library,
\texttt{libcrypto.so.1.0.0}, and were pinned to run on different cores
of the same socket, thus sharing the same last-level cache. The sender
ran in a loop, repeatedly accessing one memory location (the beginning
address of function \texttt{AES\_decrypt()}). The receiver executed
\flushreload attacks on the same memory address, by first \Flush{ing}
the memory block out of the shared LLC with an \clflush instruction
and then \Reload{ing} the memory address by accessing it directly
while measuring the access latency. The interval between \Flush and
\Reload was set to 2500 \cycles. The experiment was run for 500,000
\flushreload trials.  We then repeated this experiment with the sender
accessing an unshared address, to form a baseline.

\begin{figure}[t]
\centering
\subfigure[][with \sysname disabled]{
\includegraphics[width=0.47\linewidth]{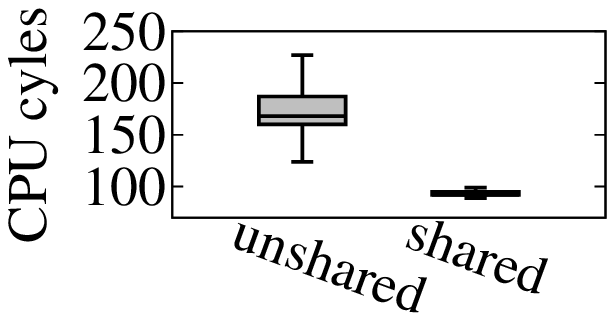}
\label{fig:flushreload_covert:linux}
}
\subfigure[][with \sysname enabled]{
\includegraphics[width=0.47\linewidth]{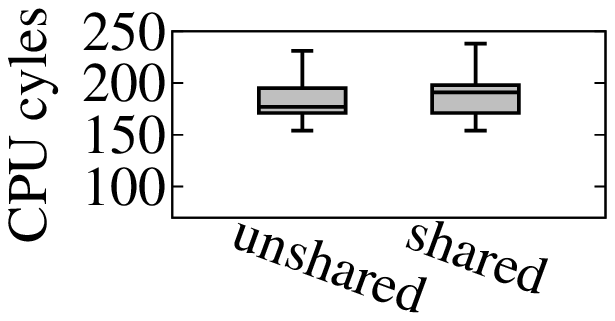}
\label{fig:flushreload_covert:cachebar}
}
\caption{\Reload timings in \flushreload attacks on an
  address shared with the victim vs.\ timings on an unshared address}
\label{fig:flushreload_covert}
\end{figure}

\figref{fig:flushreload_covert:linux} shows the results of this
experiment, when run over unmodified Linux.
The three horizontal lines forming the ``box'' in each boxplot
represents the first, second (median), and third quartiles of the
\flushreload measurements; whiskers extend to cover all points that
lie within $1.5\times$ the interquartile range.  As can be seen in
this figure, the times observed by the receiver to \Reload the shared
address were clearly separable from the times to \Reload the unshared
address, over unmodified Linux.  With \sysname enabled, however, these
measurements are no longer separable
(\figref{fig:flushreload_covert:cachebar}).

Certain corner cases are not represented in
\figref{fig:flushreload_covert}.  For example, we found it extremely
difficult to conduct experiments to capture the corner cases where
\Flush and \Reload takes place right before and after physical page
mergers, as described in \secref{sec:coa:security}. As such, we rely
on our manual inspection of the implementation in these cases to check
correctness and argue these corner cases are very difficult to exploit
in practice.

\subsubsection{\primeprobe Attacks}
\label{sec:eval:security:primeprobe}

We evaluate the effectiveness of \sysname against \primeprobe attacks
by measuring its ability to interfere with a simulated attack. In our
simulation, a process in an attacker container repeatedly performed
\primeprobe attacks on a specific cache set, while a process in a
victim container accessed data that were retrieved into the same cache
set at the rate of $\victimDemand$ accesses per attacker \primeprobe
interval.  Moreover, the cache lines available to the victim container
and attacker container, i.e., $\linesPerContainer{\victimLabel}$ and
$\linesPerContainer{\attackerLabel}$ respectively, were fixed in each
experiment.

The machine architecture on which we performed these tests had a
\cacheLineNmbr-way LLC with $\cacheLineNmbr=16$.  The values
\linesPerContainer{\attackerLabel} and
\linesPerContainer{\victimLabel} were distributed as computed in
\secref{sec:cacheabilityMgmt:security}; with this distribution, only
values in $\{4,5,6,\ldots,14\}$ were possible. In each test with fixed
\linesPerContainer{\victimLabel} and
\linesPerContainer{\attackerLabel}, we allowed the victim to place a
demand of (i.e., retrieve memory blocks to fill) $\victimDemand \in
\{0, 1, 2,...,16\}$ cache lines of the cache set undergoing the
\primeprobe attack by the attacker.  The attacker's goal was to
classify the victim's demand into one of six classes: $\classNone =
\{0\}$, $\classOne =\{1\}$, $\classFew=\{2,3,4\}$,
$\classSome=\{5,6,7,8\}$, $\classLots=\{9,10,11,12\}$, and
$\classMost=\{13,14,15,16\}$.  By asking the attacker to classify the
victim's demand into only one of six classes (versus one of 16), we
substantially simplified the attacker's job.

Also to make the attacker's job easier, we permitted the attacker to
know \linesPerContainer{\attackerLabel}; i.e., the attacker trained a
different classifier per value of \linesPerContainer{\attackerLabel},
with knowledge of the demand \victimDemand per \primeprobe trial, and
then tested against additional trial results to classify unknown
victim demands.  Specifically, after training a naive Bayes classifier
on 500,000 \primeprobe trials per $(\victimDemand, \linesPerContainer{\attackerLabel}, \linesPerContainer{\victimLabel})$ triple, we tested it on another
500,000 trials. To filter out \Probe readings due to page faults,
excessively large readings were discarded from our evaluation. The
tests without protection by \sysname yielded the confusion matrix in
\tabref{fig:confusion:disabled}, with overall accuracy of 67.5\%.  In
this table, cells with higher numbers have lighter backgrounds, and so
the best attacker would be one who achieves white cells along
the diagonal and dark-gray cells elsewhere.  As can be seen
there, classification by the attacker was very accurate for
\victimDemand falling into \classNone, \classOne, or \classLots; e.g.,
$\victimDemand=1$ resulted in a classification of \classOne with
probability of $0.80$.  Some other demands had lower accuracy, but
were almost always classified into adjacent classes; specifically,
\textit{every} class of victim demand was classified correctly or as
an adjacent class (e.g., $\victimDemand \in \classFew$ was classified
as \classOne, \classFew, or \classSome) at least 96\% of the time.

\begin{figure}[tb]
  \ExplSyntaxOn
  \fp_set:Nn \MinVal {0.0}
  \fp_set:Nn \MaxVal {1.0}
  \ExplSyntaxOff
  \begin{center}
    \subfigure[][Without \sysname]{
      \label{fig:confusion:disabled}
      {\footnotesize
      \begin{tabular}{@{\hspace{0pt}}c@{\hspace{0.25em}}rRRRRRR}
       &  & \multicolumn{6}{c}{Classification by attacker} \\
       &  & \multicolumn{1}{c}{\classNone}& \multicolumn{1}{c}{\classOne}& \multicolumn{1}{c}{\classFew}& \multicolumn{1}{c}{\classSome}& \multicolumn{1}{c}{\classLots}& \multicolumn{1}{c}{\classMost}\\\cline{3-8}\\[-2.485ex]
        \multirow{6}{*}{\rotatebox[origin=c]{90}{\parbox{4em}{\centering Victim\\[-3pt] demand \victimDemand}}} 
       &  \multicolumn{1}{r|}{\classNone}& .96& .04& .00& .00& .00& .00 \\
       & \multicolumn{1}{r|}{\classOne}& .01& .80& .19& .01& .00& .00 \\
       &  \multicolumn{1}{r|}{\classFew}& .00& .16& .50& .30& .04& .00 \\
       &  \multicolumn{1}{r|}{\classSome}& .00& .00& .07& .54& .34& .04 \\
       &  \multicolumn{1}{r|}{\classLots}& .00& .00& .00& .03& .84& .13 \\
       &  \multicolumn{1}{r|}{\classMost}& .00& .00& .00& .03& .56& .41 \\
        \multicolumn{7}{c}{~} \\
      \end{tabular}
      }
    } \\
    \subfigure[][With \sysname]{
      \label{fig:confusion:enabled}
      {\footnotesize
       \begin{tabular}{@{\hspace{0pt}}c@{\hspace{0.25em}}rRRRRRR}
       	&  & \multicolumn{6}{c}{Classification by attacker} \\
       	&  & \multicolumn{1}{c}{\classNone}& \multicolumn{1}{c}{\classOne}& \multicolumn{1}{c}{\classFew}& \multicolumn{1}{c}{\classSome}& \multicolumn{1}{c}{\classLots}& \multicolumn{1}{c}{\classMost}\\\cline{3-8}\\[-2.485ex]
       	\multirow{6}{*}{\rotatebox[origin=c]{90}{\parbox{4em}{\centering Victim\\[-3pt] demand \victimDemand}}} 
           & \multicolumn{1}{r|}{\classNone}& .33& .16& .26& .18& .04& .02 \\
           & \multicolumn{1}{r|}{\classOne}& .16& .36& .19& .19& .06& .04 \\
           & \multicolumn{1}{r|}{\classFew}& .13& .14& .40& .19& .09& .05 \\
           & \multicolumn{1}{r|}{\classSome}& .09& .10& .16& .37& .20& .07 \\
           & \multicolumn{1}{r|}{\classLots}& .08& .06& .10& .16& .46& .13 \\
           & \multicolumn{1}{r|}{\classMost}& .10& .07& .18& .18& .18& .29 \\
       	\multicolumn{8}{c}{~} \\
       \end{tabular}
       }
    }
\end{center}
\caption{Confusion matrix of naive Bayes classifier}
\label{fig:confusion}
\end{figure}

In contrast, \figref{fig:confusion:enabled} shows the confusion matrix
for a naive Bayes classifier trained and tested using \primeprobe
trials conducted with \sysname enabled.  Specifically, these values
were calculated using
\begin{align*}
&\cprob{\big}{}{\classifiedRV = \classId}{\victimDemand\in\classIdAlt} \\
&= \sum_{4 \le \attackerLineNmbr, \victimLineNmbr \le 14}
  \left(\begin{array}{r}
    \cprob{\Big}{}{\classifiedRV = \classId}{\begin{array}{r}\victimDemand\in\classIdAlt \wedge \victimLineNmbrRV = \victimLineNmbr \\
  \wedge~\attackerLineNmbrRV = \attackerLineNmbr\end{array}} \\
  \cdot~\prob{}{\attackerLineNmbrRV = \attackerLineNmbr}
  \cdot \prob{}{\victimLineNmbrRV = \victimLineNmbr}
  \end{array}\right)
\end{align*}
where \classifiedRV denotes the classification obtained by the
adversary using the naive Bayes classifier; $\classId, \classIdAlt \in
\{\classNone$, \classOne, \classFew, \classSome, \classLots,
$\classMost\}$; and $\prob{}{\attackerLineNmbrRV = \attackerLineNmbr}$
and $\prob{}{\victimLineNmbrRV = \victimLineNmbr}$ are calculated as
described in \secref{sec:cacheabilityMgmt:security}.  The factor
$\cprob{\big}{}{\classifiedRV = \classId}{\victimDemand\in\classIdAlt
  \wedge \victimLineNmbrRV = \victimLineNmbr \wedge
  \attackerLineNmbrRV = \attackerLineNmbr}$ was measured empirically.
Though space limits preclude reporting the full class confusion matrix
for each \victimLineNmbr, \attackerLineNmbr pair, the accuracy of the
naive Bayes classifier per \victimLineNmbr, \attackerLineNmbr pair,
averaged over all classes \classId, is shown in \figref{fig:accuracy}.
As in \figref{fig:confusion}, cells with larger values in
\figref{fig:accuracy} are more lightly colored, though in this case,
the diagonal has no particular significance.  Rather, we would expect
that when the attacker and victim are each limited to fewer lines in
the cache set (i.e., small values of \attackerLineNmbr and
\victimLineNmbr, in the upper left-hand corner of
\figref{fig:accuracy}) the accuracy of the attacker will suffer,
whereas when the attacker and victim are permitted to use more lines
of the cache (i.e., in the lower right-hand corner) the attacker's
accuracy would improve.  \figref{fig:accuracy} supports these general
trends.

\begin{figure}[t]
\begin{center}
\ExplSyntaxOn
\fp_set:Nn \MinVal {0.17}
\fp_set:Nn \MaxVal {0.68}
\ExplSyntaxOff
{\tiny
\begin{tabular}{@{\hspace{0pt}}c@{\hspace{0.5em}}r|TTTTTTTTTTT}
& \multicolumn{1}{r}{~} & \multicolumn{11}{c}{$\linesPerContainer{\victimLabel}$} \\
& \multicolumn{1}{r}{~} & \multicolumn{1}{c}{4}& \multicolumn{1}{c}{5}& \multicolumn{1}{c}{6}& \multicolumn{1}{c}{7}& \multicolumn{1}{c}{8}& \multicolumn{1}{c}{9}& \multicolumn{1}{c}{10}& \multicolumn{1}{c}{11}& \multicolumn{1}{c}{12}& \multicolumn{1}{c}{13}& \multicolumn{1}{c}{14} \\\cline{3-13}\\[-2.485ex]
\parbox[t]{1mm}{\multirow{11}{*}{\rotatebox[origin=c]{90}{$\linesPerContainer{\attackerLabel}$}}} 
& 4 & .18& .17& .17& .17& .17& .17& .17& .17& .36& .22& .33 \\
& 5 & .19& .17& .30& .32& .27& .27& .20& .26& .33& .46& .39 \\
& 6 & .17& .31& .24& .18& .21& .17& .20& .27& .43& .39& .41 \\
& 7 & .17& .33& .22& .22& .19& .31& .33& .33& .46& .48& .54 \\
& 8 & .33& .35& .32& .23& .43& .37& .43& .42& .32& .38& .49 \\
& 9 & .20& .26& .31& .28& .44& .38& .34& .34& .46& .39& .56 \\
& 10& .41& .31& .27& .35& .50& .55& .53& .31& .53& .50& .62 \\
& 11& .45& .45& .40& .45& .47& .54& .54& .57& .67& .50& .50 \\
& 12& .55& .50& .59& .63& .49& .48& .54& .49& .56& .58& .57 \\
& 13& .55& .53& .68& .68& .54& .65& .52& .56& .57& .66& .66 \\
& 14& .53& .56& .45& .65& .46& .62& .48& .68& .55& .57& .53 \\
\end{tabular}
}
\end{center}
\caption{Accuracy per values of \victimLineNmbr and \attackerLineNmbr}
\label{fig:accuracy}
\end{figure}

Returning to \figref{fig:confusion:enabled}, we see that \sysname
substantially degrades the adversary's classification accuracy, which
overall is only 33\%.  Moreover, the adversary is not only
wrong more often, but is also often ``more wrong'' in those cases.
That is, whereas in \figref{fig:confusion:disabled} shows that each
class of victim demand was classified as that demand or an adjacent
demand at least 96\% of the time, this property no longer holds true
in \figref{fig:confusion:enabled}.  Indeed, the attacker's
\textit{best} case in this regard is classifying victim demand
\classLots, which it classifies as \classSome, \classLots, or
\classMost 75\% of the time.  In the case of a victim demand of
\classMost, this number is only 47\%.

\subsection{Performance Evaluation}
\label{sec:eval:perf}

In this section we describe tests we have run to evaluate the
performance impact of \sysname relative to an unmodified Linux
kernel. As mentioned previously, we are motivated by side-channel
prevention in PaaS clouds, and so we focused our performance
evaluation on typical PaaS applications (primarily web servers
supporting various language runtimes) running together with \sysname.
For the sake of space, we defer our discussion on typical PaaS
applications to \appref{sec:paas}.
Also, here we report only throughput and response-time measurements;
other experiments to shed light on \sysname's memory savings over
prohibiting cross-container memory sharing in Linux as
an alternative to copy-on-access, can be found in \appref{sec:memory}.

\begin{figure*}[t]
\centering
\subfigure[][4 webservers]{
\label{fig:dynamic:4}
\includegraphics[width=0.663\columnwidth]{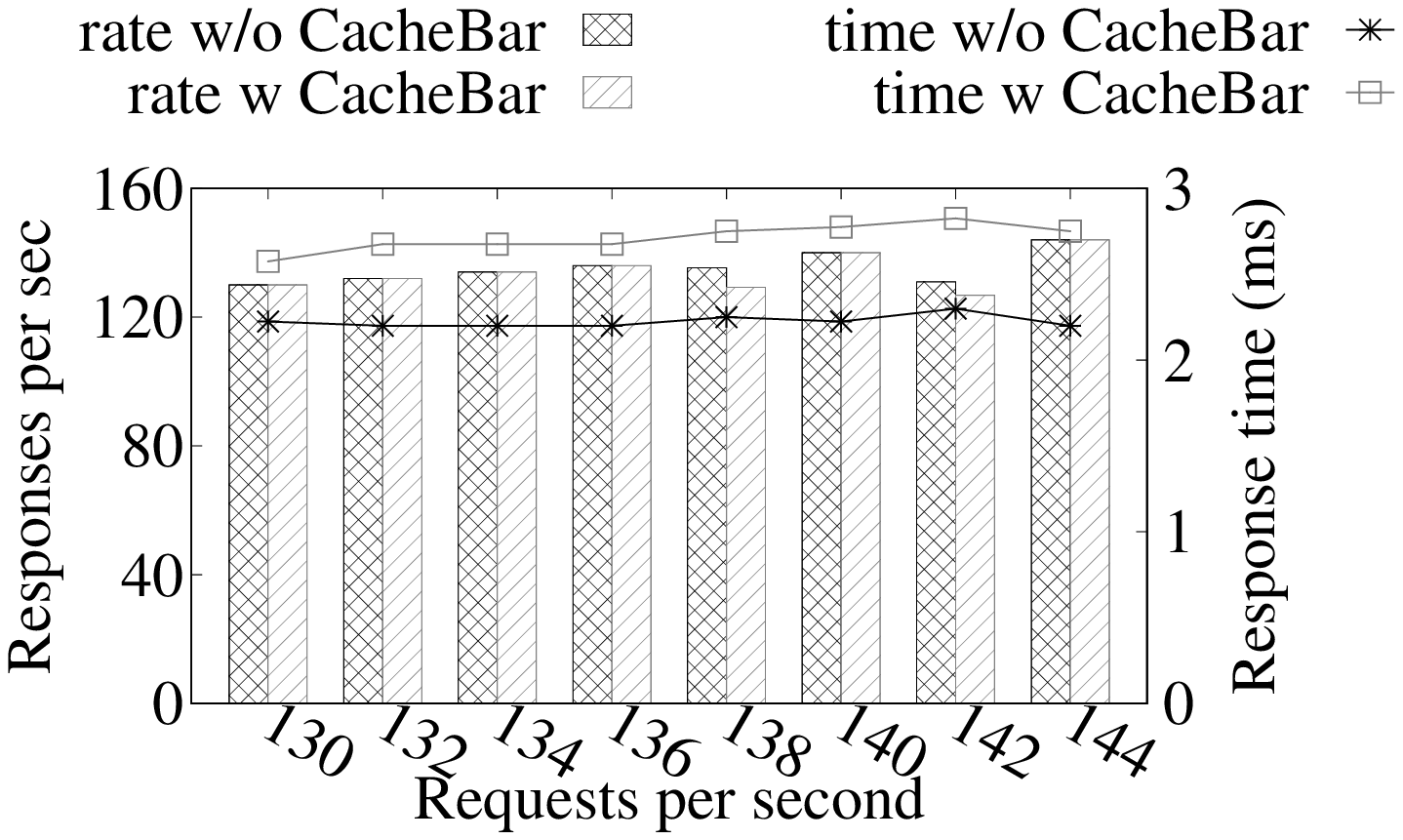}
}
\subfigure[][16 webservers]{
\label{fig:dynamic:16}
\includegraphics[width=0.663\columnwidth]{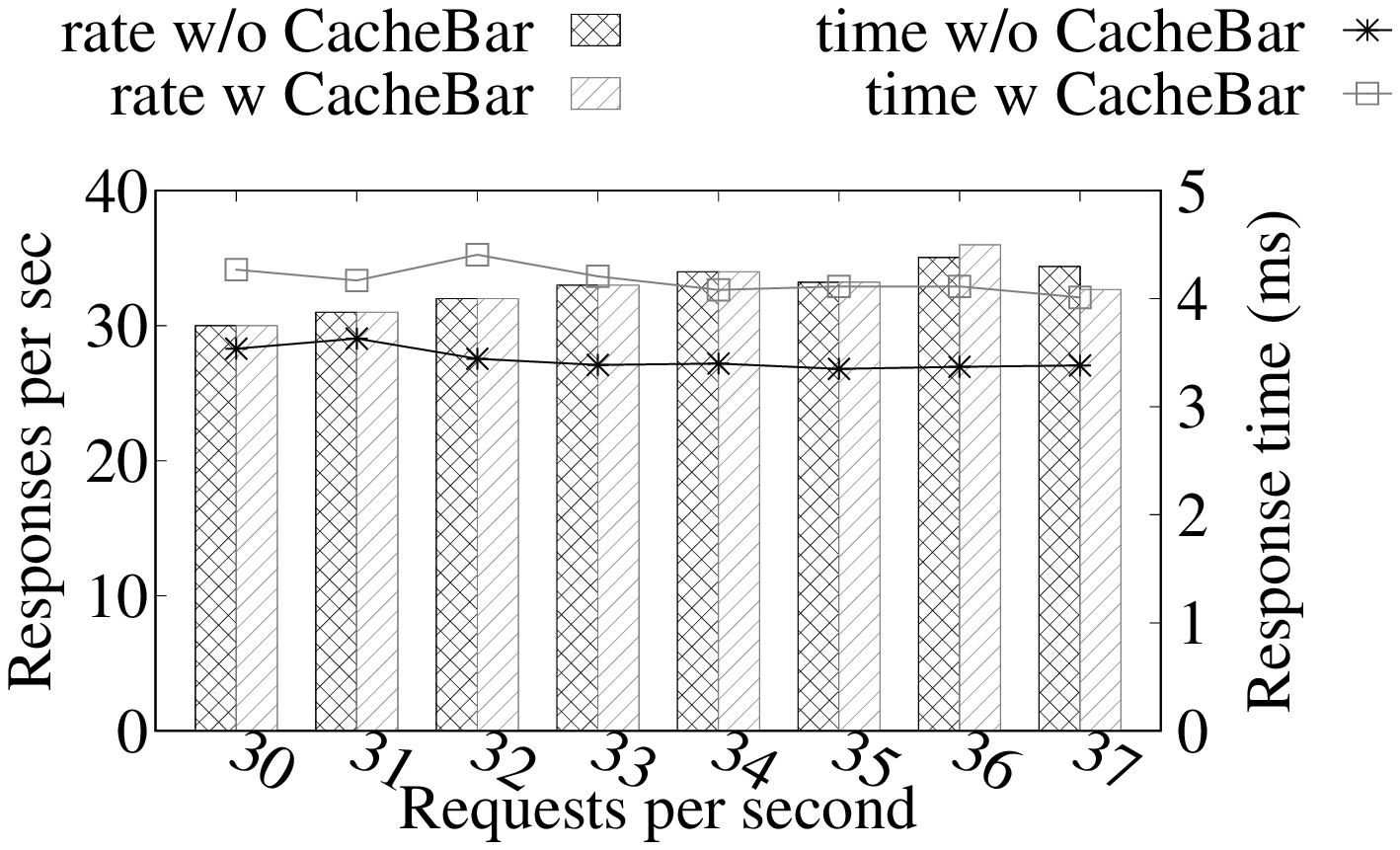}
}
\subfigure[][different numbers of webservers]{
\label{fig:dynamic:all}
\includegraphics[width=0.663\columnwidth]{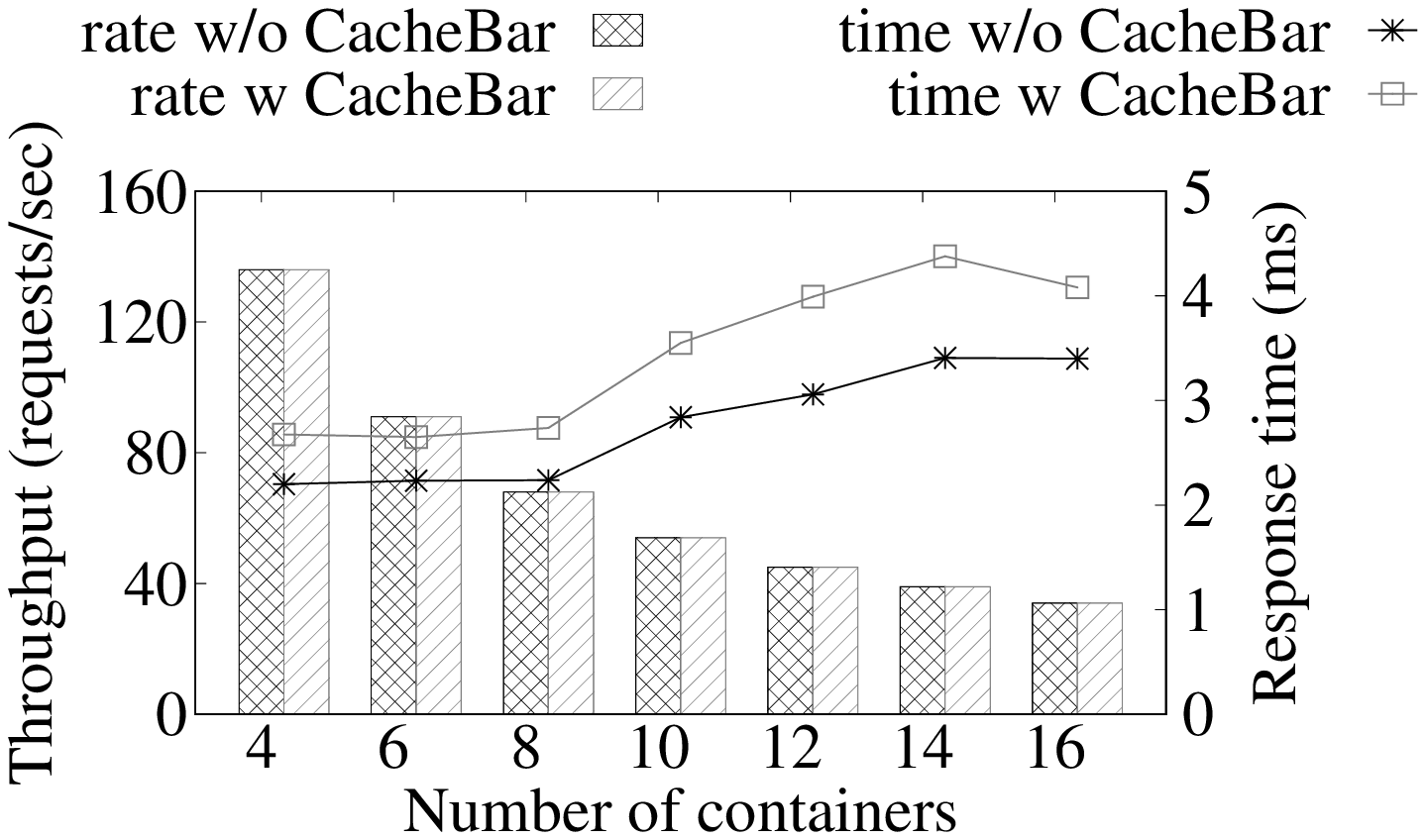}
}
\caption{Average throughput and response time per \apache+\phpfpm web
  server, each in a separate container}
\label{fig:dynamic:throughput}
\end{figure*}

Our experiments to explored \sysname's
performance impact (1) as a function of the number of container (and
webserver) instances; (2) for different combinations of webserver and
application language; (3) for complex workloads characteristic of a
social networking website; and (4) for media-streaming workloads.

\bheading{Webserver performance}
In the first experiments, each container ran an \apache version 2.4.7
web server with \phpfpm and \ssl enabled.  We set up one client per
server using \autobench; clients were spread across four computers,
each with the same networking capabilities as the (one) server
computer (not to mention more cores and memory than the server
computer), to ensure that any bottlenecks were on the server machine.
Each client repeatedly requested a web page and recorded its
achievable throughputs and response times at those throughput rates.
The content returned to each client request was the 86\kbytes output
of \phpinfo.

\figref{fig:dynamic:throughput} shows the throughputs and response
times when clients sent requests using \ssl without reusing
connections.  In particular, \figref{fig:dynamic:4} shows the achieved
response rates (left axis) and response times (right axis), averaged
over all containers, as a function of offered load when there were
four containers (and so four web servers).  Bars depict average
response rates running over unmodified Linux (``rate w/o \sysname'') or \sysname
(``rate w \sysname''), and lines depict average response times running over
unmodified Linux (``time w/o \sysname'') or \sysname (``time w \sysname'').
\figref{fig:dynamic:16} shows the same information for 16 containers.
As can be seen in these figures, the throughput impact of \sysname was
minimal, while the response time increased by around $20\%$.
\figref{fig:dynamic:all} shows this information in another way, with
the number of containers (and hence servers) increasing along the
horizontal-axis.  In \figref{fig:dynamic:all}, each bar represents the
largest request rate at which the responses could keep up.

\begin{figure*}[th]
\begin{minipage}[b]{0.35\textwidth}
\includegraphics[width=\textwidth]{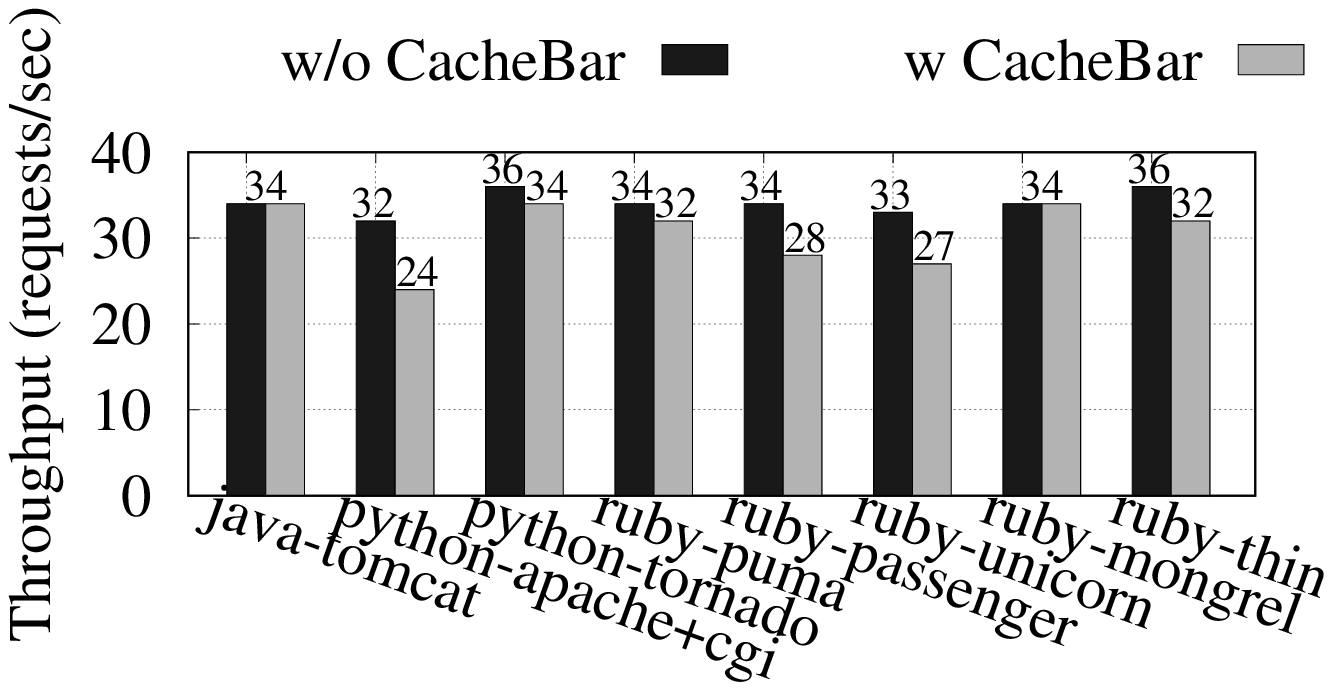}
\caption{Throughput per webserver/language}
\label{fig:diffservers}
\end{minipage}
\begin{minipage}[b]{0.35\textwidth}
\includegraphics[width=\textwidth]{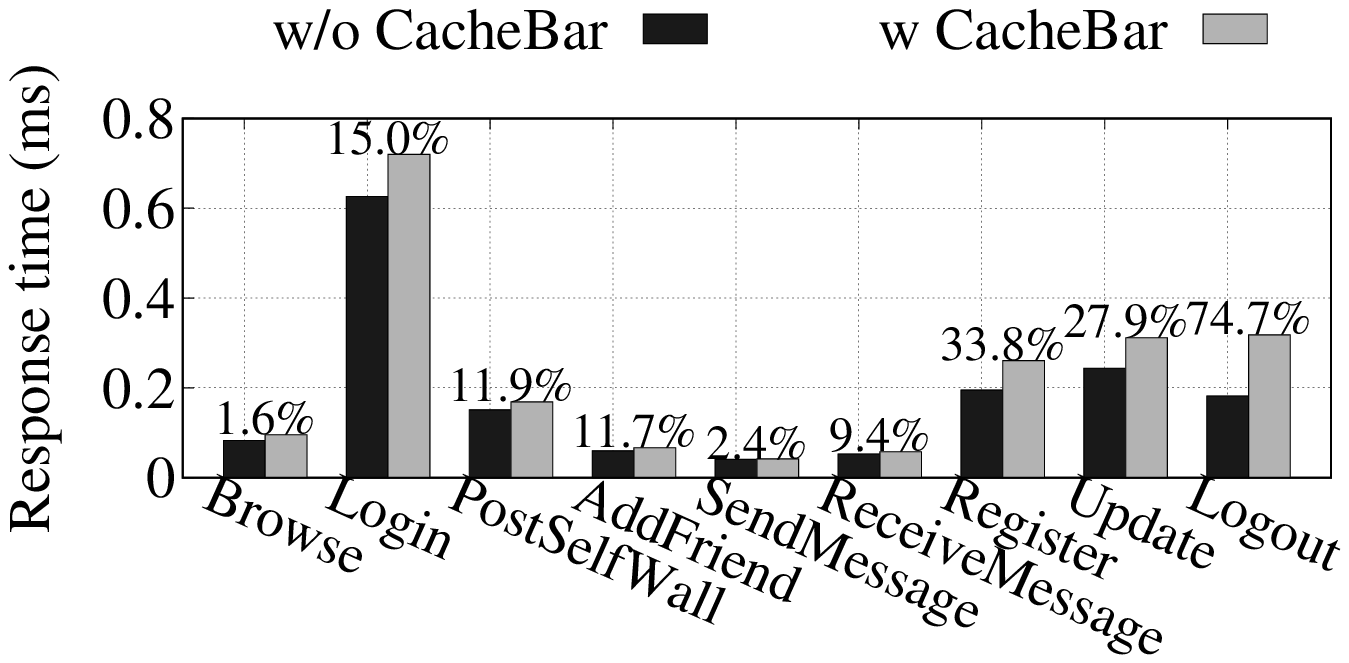}
\caption{Response times per operation}
\label{fig:diffops}
\end{minipage}
\begin{minipage}[b]{0.25\textwidth}
\includegraphics[width=\textwidth]{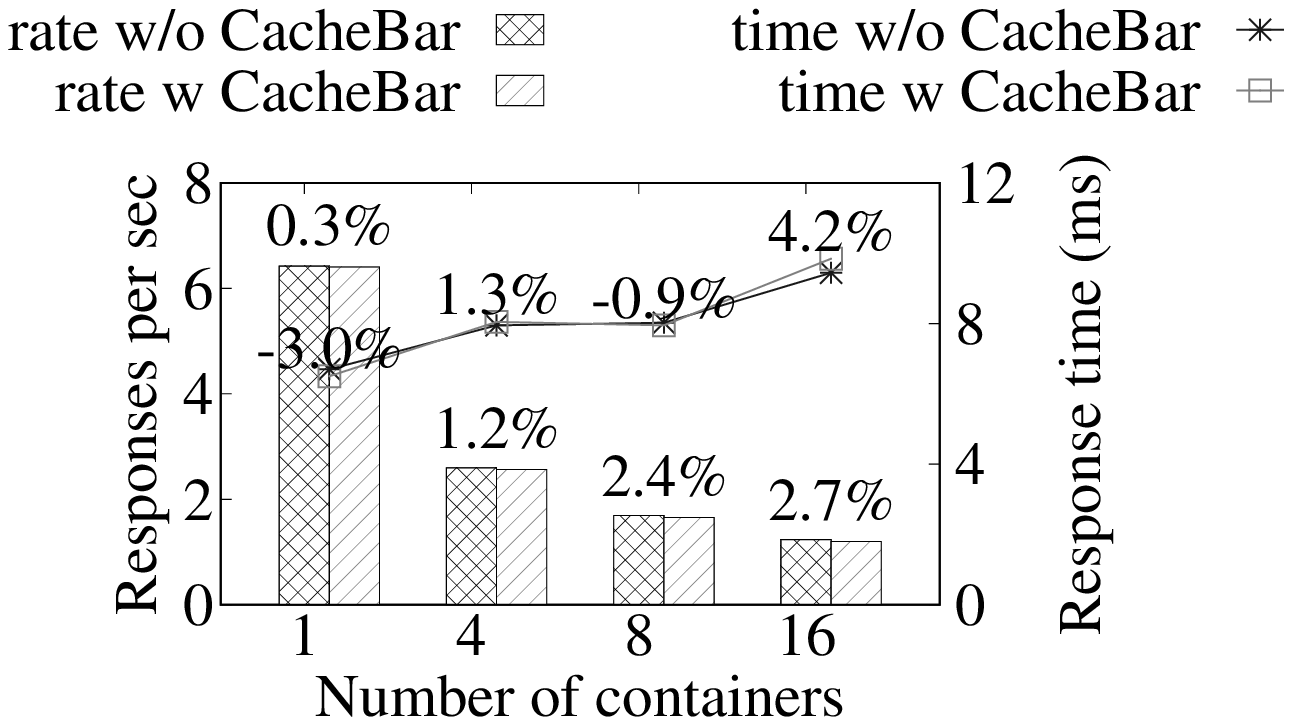}
\caption{Media streaming}
\label{fig:mediastream}
\end{minipage}
\end{figure*}

\bheading{Webserver+language combinations}
Next, we selected other common webserver+app-language combinations
(again, see \tabref{tbl:survey}), namely Java over a \tomcat web
server, Python over \apache{}+\cgi, Python over \tornado, and Ruby
over \puma.  For each configuration, we instantiated 16 containers and
set each up to dynamically generate $80\kbytes$ random strings for
clients.  We also did tests using another four web servers running the
same Ruby application, namely \passenger, \unicorn, \thin, and
\mongrel.  \figref{fig:diffservers} shows the throughput that resulted
in each case, over Linux  and over \sysname.  As shown there, the throughput overheads were
modest for most of the server+language combinations that we
considered.  The worst case was Python over \apache{}+\cgi, which
suffered a throughput degradation with \sysname of 25\%; other
degradations were much more modest.

\bheading{Impact on a more complex workload}
To test for effects on more complex workloads, we used the webserver
instance in \cloudsuite\cite{paper:cloudsuite} that implements a social community website
written in \php over \nginx on our \sysname-protected machine.  This
implementation queries a \mysql database and caches results using
\memcached; in keeping with PaaS architectures (see
\appref{sec:paas}), the database and \memcached server were
implemented on another machine without \sysname protection, since
tenants cannot typically execute directly on these machines.  We used
the \faban tool to generate a mix of requests to the
webserver, including \browse (7.9\%), \login (7.5\%), \post (24.9\%),
\addfriend (7.3\%), \sending (44.0\%), \register (0.8\%), and \logout
(7.5\%).  In addition, a background activity happened on the webserver
every 10\secs, which was either \receiving or \update with equal
likelihood.  \figref{fig:diffops} shows that the responsiveness of the
various common operations suffered little with \sysname, suffering
between 2\% and 15\% overhead.  Three operations (\register, \update,
and \logout) suffered greater than 25\% overhead, but these operations
were rare in the \faban workload (and presumably in practice).

\bheading{Media streaming in \cloudsuite}
In addition to the webserver benchmark setup used above, \cloudsuite
offers a media streaming server running over \nginx that serves
3.1\gbytes static video files at different levels of quality.  We set
up a client process per server to issue a mix of requests for videos
at different quality levels and, through a binary search, to find the
peak request rate the server can sustain while keeping the failure
rate below a threshold.  \figref{fig:mediastream} shows that \sysname
affected this application least of all, in both throughput and
response time.

\section{Conclusion}
\label{sec:conclude}

Side-channel attacks via the LLC are becoming increasingly efficient
and powerful.  To counter this growing threat, we have presented the
design of two techniques to defend against these attacks, namely (i)
copy-on-access for physical pages shared among multiple security
domains, to interfere with \flushreload attacks, and (ii) cacheability
management for pages to limit the number of cache lines per cache set
that an adversary can occupy simultaneously, to mitigate \primeprobe
attacks.  We described the implementation of these techniques in a
memory-management subsystem called \sysname for Linux, to interfere
with LLC-based side-channel attacks across containers.  We confirmed
that our design mitigates side-channel attacks through formal analysis
of both copy-on-access (using model checking) and cacheability
management (through probabilistic analysis), as well as using
empirical evaluations.  Our experiments also confirmed that the
overheads of our approach are modest for PaaS workloads, e.g.,
imposing a virtually unnoticeable cost on server throughputs.

\paragraph{Acknowledgments}
This work was supported in part by NSF grant 1330599.

{\footnotesize \bibliographystyle{acm}
\bibliography{main}}

\begin{thebibliography}{10}

\bibitem{arcangeli2009increasing}
{\sc Arcangeli, A., Eidus, I., and Wright, C.}
\newblock Increasing memory density by using {KSM}.
\newblock In {\em Linux Symposium\/} (2009), pp.~19--28.

\bibitem{azar2014colocation}
{\sc Azar, Y., Kamara, S., Menache, I., Raykova, M., and Shepard, B.}
\newblock Co-location-resistant clouds.
\newblock In {\em 6th ACM Cloud Computing Security Workshop\/} (2014),
  pp.~9--20.

\bibitem{coppens2009practical}
{\sc Coppens, B., Verbauwhede, I., Bosschere, K.~D., and Sutter, B.~D.}
\newblock Practical mitigations for timing-based side-channel attacks on modern
  x86 processors.
\newblock In {\em 30th IEEE Symposium on Security and Privacy\/} (2009),
  pp.~45--60.

\bibitem{crane2015thwarting}
{\sc Crane, S., Homescu, A., Brunthaler, S., Larsen, P., and Franz, M.}
\newblock Thwarting cache side-channel attacks through dynamic software
  diversity.
\newblock In {\em 2015 ISOC Network and Distributed System Security
  Symposium\/} (2015).

\bibitem{elizaveta2001emd}
{\sc Elizaveta, L., and Bickel, P.}
\newblock The earth mover's distance is the {Mallows} distance.
\newblock In {\em 8th International Conference on Computer Vision\/} (2001),
  pp.~251--256.

\bibitem{fabrega1995copy}
{\sc F{\'a}brega, F. J.~T., and Guttman, J.~D.}
\newblock Copy on write.
\newblock Citeseer, 1995.

\bibitem{paper:cloudsuite}
{\sc Ferdman, M., Adileh, A., Kocberber, O., Volos, S., Alisafaee, M., Jevdjic,
  D., Kaynak, C., Popescu, A.~D., Ailamaki, A., and Falsafi, B.}
\newblock Clearing the clouds: a study of emerging scale-out workloads on
  modern hardware.
\newblock In {\em 17th International Conference on Architectural Support for
  Programming Languages and Operating Systems\/} (2012), pp.~37--48.

\bibitem{gruss:2015:FF}
{\sc Gruss, D., Maurice, C., and Wagner, K.}
\newblock {Flush+Flush}: A stealthier last-level cache attack.
\newblock {\em CoRR abs/1511.04594\/} (Nov. 2015).

\bibitem{gullasch2011games}
{\sc Gullasch, D., Bangerter, E., and Krenn, S.}
\newblock Cache games -- bringing access-based cache attacks on {AES} to
  practice.
\newblock In {\em 32nd IEEE Symposium on Security and Privacy\/} (May 2011),
  pp.~490--505.

\bibitem{han2013security}
{\sc Han, Y., Alpcan, T., Chan, J., and Leckie, C.}
\newblock Security games for virtual machine allocation in cloud computing.
\newblock In {\em 4th International Conference Decision and Game Theory for
  Security}, vol.~8252 of {\em LNCS}. Nov. 2013, pp.~99--118.

\bibitem{guide2010intel}
{\sc Intel}.
\newblock {\em Intel{\textregistered} 64 and IA-32 Architectures Software
  Developer’s Manual}, 2010.

\bibitem{irazoqui2015shared}
{\sc Irazoqui, G., Eisenbarth, T., and Sunar, B.}
\newblock {S\$A}: A shared cache attack that works across cores and defies {VM}
  sandboxing---and its application to {AES}.
\newblock In {\em 36th IEEE Symposium on Security and Privacy\/} (May 2015).

\bibitem{keramidas2008non}
{\sc Keramidas, G., Antonopoulos, A., Serpanos, D.~N., and Kaxiras, S.}
\newblock Non deterministic caches: A simple and effective defense against side
  channel attacks.
\newblock {\em Design Automation for Embedded Systems 12}, 3 (2008), 221--230.

\bibitem{kim2012stealthmem}
{\sc Kim, T., Peinado, M., and Mainar-Ruiz, G.}
\newblock {STEALTHMEM}: System-level protection against cache-based side
  channel attacks in the cloud.
\newblock In {\em USENIX Security Symposium\/} (2012), pp.~189--204.

\bibitem{kong2008deconstructing}
{\sc Kong, J., Aciicmez, O., Seifert, J.~P., and Zhou, H.}
\newblock Deconstructing new cache designs for thwarting software cache-based
  side channel attacks.
\newblock In {\em 2nd ACM Workshop on Computer Security Architectures\/}
  (2008), pp.~25--34.

\bibitem{konighofer2008fast}
{\sc K{\"o}nighofer, R.}
\newblock A fast and cache-timing resistant implementation of the aes.
\newblock In {\em Topics in Cryptology--CT-RSA 2008}. 2008, pp.~187--202.

\bibitem{li2012improving}
{\sc Li, M., Zhang, Y., Bai, K., Zang, W., Yu, M., and He, X.}
\newblock Improving cloud survivability through dependency based virtual
  machine placement.
\newblock In {\em International Conference on Security and Cryptography\/}
  (July 2012), pp.~321--326.

\bibitem{li2014stopwatch}
{\sc Li, P., Gao, D., and Reiter, M.~K.}
\newblock {StopWatch}: A cloud architecture for timing channel mitigation.
\newblock {\em ACM Transations on Information and System Security 17}, 2 (Nov.
  2014).

\bibitem{CATalyst}
{\sc Liu, F., Ge, Q., Yarom, Y., Mckeen, F., Rozas, C., Heiser, G., and Lee,
  R.~B.}
\newblock Catalyst: Defeating last-level cache side channel attacks in cloud
  computing.
\newblock In {\em 22nd IEEE Symposium on High Performance Computer
  Architecture\/} (Mar. 2016).

\bibitem{Liu:2014:RFC}
{\sc Liu, F., and Lee, R.~B.}
\newblock Random fill cache architecture.
\newblock In {\em 47th Annual IEEE/ACM International Symposium on
  Microarchitecture\/} (2014), IEEE Computer Society, pp.~203--215.

\bibitem{liu2015practical}
{\sc Liu, F., Yarom, Y., Ge, Q., Heiser, G., and Lee, R.~B.}
\newblock Last-level cache side-channel attacks are practical.
\newblock In {\em 36th IEEE Symposium on Security and Privacy\/} (May 2015).

\bibitem{mallows1972emd}
{\sc Mallows, C.~L.}
\newblock A note on asymptotic joint normality.
\newblock {\em Annals of Mathematical Statistics 43}, 2 (1972), 508--515.

\bibitem{osvik2006cache}
{\sc Osvik, D.~A., Shamir, A., and Tromer, E.}
\newblock Cache attacks and countermeasures: the case of {AES}.
\newblock In {\em Topics in Cryptology--CT-RSA 2006}. Springer, 2006,
  pp.~1--20.

\bibitem{page2005partitioned}
{\sc Page, D.}
\newblock Partitioned cache architecture as a side-channel defence mechanism.
\newblock Tech. Rep. 2005/280, IACR Cryptology ePrint Archive, 2005.

\bibitem{raikin2014tracking}
{\sc Raikin, S., Sager, J.~D., Sperber, Z., Krimer, E., Lempel, O., Shwartsman,
  S., Yoaz, A., and Golz, O.}
\newblock Tracking mechanism coupled to retirement in reorder buffer for
  indicating sharing logical registers of physical register in record indexed
  by logical register, Dec.~16 2014.
\newblock US Patent 8,914,617.

\bibitem{raj2009resource}
{\sc Raj, H., Nathuji, R., Singh, A., and England, P.}
\newblock Resource management for isolation enhanced cloud services.
\newblock In {\em 2009 ACM Workshop on Cloud Computing Security\/} (2009),
  pp.~77--84.

\bibitem{shi2011limiting}
{\sc Shi, J., Song, X., Chen, H., and Zang, B.}
\newblock Limiting cache-based side-channel in multi-tenant cloud using dynamic
  page coloring.
\newblock In {\em Workshops of the 41st IEEE/IFIP International Conference on
  Dependable Systems and Networks\/} (2011), pp.~194--199.

\bibitem{stefan2013eliminating}
{\sc Stefan, D., Buiras, P., Yang, E.~Z., Levy, A., Terei, D., Russo, A., and
  Mazi{\`e}res, D.}
\newblock Eliminating cache-based timing attacks with instruction-based
  scheduling.
\newblock In {\em Computer Security--ESORICS 2013}. 2013, pp.~718--735.

\bibitem{Svenningsson:2009:SVS}
{\sc Svenningsson, J., and Sands, D.}
\newblock Specification and verification of side channel declassification.
\newblock In {\em 6th International Conference on Formal Aspects in Security
  and Trust\/} (2010), Springer-Verlag, pp.~111--125.

\bibitem{tromer2010efficient}
{\sc Tromer, E., Osvik, D.~A., and Shamir, A.}
\newblock Efficient cache attacks on {AES}, and countermeasures.
\newblock {\em Journal of Cryptology 23}, 1 (2010), 37--71.

\bibitem{varadarajan2014scheduler}
{\sc Varadarajan, V., Ristenpart, T., and Swift, M.}
\newblock Scheduler-based defenses against cross-{VM} side channels.
\newblock In {\em 23rd USENIX Security Symposium\/} (Aug. 2014).

\bibitem{vattikonda2011timers}
{\sc Vattikonda, B.~C., Das, S., and Shacham, H.}
\newblock Eliminating fine grained timers in {Xen}.
\newblock In {\em 3rd ACM Cloud Computing Security Workshop\/} (Oct. 2011),
  pp.~41--46.

\bibitem{wang2008novel}
{\sc Wang, Z., and Lee, R.~B.}
\newblock A novel cache architecture with enhanced performance and security.
\newblock In {\em 41st IEEE/ACM International Symposium on Microarchitecture\/}
  (2008), pp.~83--93.

\bibitem{wray1991analysis}
{\sc Wray, J.~C.}
\newblock An analysis of covert timing channels.
\newblock In {\em 1991 IEEE Symposium on Security and Privacy\/} (1991),
  pp.~2--7.

\bibitem{yarom2014flush}
{\sc Yarom, Y., and Falkner, K.~E.}
\newblock {FLUSH+RELOAD}: A high resolution, low noise, {L3} cache side-channel
  attack.
\newblock In {\em 23rd USENIX Security Symposium\/} (2014), pp.~719--732.

\bibitem{zhang2012cross}
{\sc Zhang, Y., Juels, A., Reiter, M.~K., and Ristenpart, T.}
\newblock Cross-{VM} side channels and their use to extract private keys.
\newblock In {\em ACM Conference on Computer \& Communications Security\/}
  (2012), pp.~305--316.

\bibitem{zhang2014cross}
{\sc Zhang, Y., Juels, A., Reiter, M.~K., and Ristenpart, T.}
\newblock Cross-tenant side-channel attacks in {PaaS} clouds.
\newblock In {\em ACM Conference on Computer \& Communications Security\/}
  (2014), pp.~990--1003.

\bibitem{zhang2012incentive}
{\sc Zhang, Y., Li, M., Bai, K., Yu, M., and Zang, W.}
\newblock Incentive compatible moving target defense against {VM}-colocation
  attacks in clouds.
\newblock In {\em 27th IFIP Information Security and Privacy Conference},
  vol.~376 of {\em IFIP Advances in Information and Communication Technology}.
  June 2012, pp.~388--399.

\bibitem{zhang2013duppel}
{\sc Zhang, Y., and Reiter, M.~K.}
\newblock D{\"u}ppel: Retrofitting commodity operating systems to mitigate
  cache side channels in the cloud.
\newblock In {\em 2013 ACM Conference on Computer \& Communications Security\/}
  (2013), pp.~827--838.

\end{thebibliography}

\appendix

\section{Platform-as-a-Service Cloud}
\label{sec:paas}

Platform-as-a-Service (PaaS) cloud is a model of cloud computing which
enables users to develop and deploy web applications without
installing and managing the required in-house hardware and software. A
typical PaaS cloud supports multiple programming languages and
facilitates the integration of a variety of application middleware
(e.g., data analytics, messaging, and load balancing) and databases
(e.g., memcache, SQL). For instance, the popular Heroku
(\url{heroku.com}) service supports more than ten programming
languages such as Ruby, Node.js, Python, and Java, PHP, Go, Perl, C,
Erlang, Scala, and Clojure, and provides integration of application
middleware as add-ons to facilitate data storage, mobile integration,
monitoring and logging, and other types of application development.
Customers of PaaS usually need to develop their applications on local
machines, and then upload source code to the PaaS system via for
testing and deployment.  In some cases, they are allowed to \ssh onto
the remote machine and perform necessary configuration and debugging.

In order to increase server utilization and reduce cost, most public
PaaS offerings are multi-tenant, which serve multiple customers on the
same (virtual) machine.  Tenants sharing the same operating system are
typically isolated using Linux containers. While a web application may contain web
servers, programming language runtimes, databases, and a set of
middleware that enrich its functionality, in all PaaS clouds we have
studied, language runtimes and web servers are located on different
servers from databases and middleware; web servers controlled by
different tenants may share the same OS, however. Because users of
PaaS clouds do not have the permission to execute arbitrary code on
databases and middleware that are typically shared by multiple
tenants, the targets of the \primeprobe and \flushreload side-channel
attacks we consider in this paper are primarily web servers that
supports various language runtimes, which may be co-located with the
adversary-controlled malicious web servers on which arbitrary code can
be executed. We conducted a survey to understand the web servers that
are used in major PaaS clouds, and the programming languages they
support. The results are shown in \tabref{tbl:survey}.

\begin{table*}
\centering
{\footnotesize
\renewcommand{\arraystretch}{1.2}
\begin{tabular}{p{0.18\textwidth}p{0.40\textwidth}p{0.36\textwidth}}
\toprule
PaaS cloud & Supported server engines & Supported programming languages \\
\midrule
\parbox[t]{0.18\textwidth}{\appfog \\[-3pt] {\tiny\url{www.appfog.com}}}
& Apache Tomcat, Apache HTTP, Nginx, Microsoft IIS
& Java, Python, PHP, Node.js, Ruby and Go \\
\parbox[t]{0.18\textwidth}{\azure \\[-3pt] {\tiny\url{azure.microsoft.com}}}
& Apache Tomcat, Jetty, Apache HTTP, Nginx, GlassFish, Wildfly, Jetty, Microsoft IIS
& Java, Python, PHP, Node.js, and ASP.NET \\
\parbox[t]{0.18\textwidth}{\dotcloud \\[-3pt] {\tiny\url{www.dotcloud.com}}}
& Apache Tomcat, Tornado, PHP built-in webserver
& Java, Python, PHP, Node.js, Ruby and Go\\[8pt]
\parbox[t]{0.18\textwidth}{\elasticbeanstalk \\[-3pt] {\tiny\url{aws.amazon.com/elasticbeanstalk}}}
& Apache Tomcat, Apache HTTP or Nginx, Passenger or Puma, Microsoft IIS
& Java, Python, PHP, Node.js, Ruby, Go, ASP.NET, and others\\
\parbox[t]{0.18\textwidth}{\engineyard \\[-3pt] {\tiny\url{www.engineyard.com}}}
& Nginx, Rack, Passenger, Puma, Unicorn, Trinidad
& Java, PHP, Node.js, and Ruby\\[8pt]
\parbox[t]{0.18\textwidth}{Google Cloud \\[-3pt] {\tiny\url{cloud.google.com/appengine}}}
& JBoss, Wildfly, Apache Tomcat, Apache HTTP, Nginx, Zend Server, Passenger, Mongrel, Thin, Microsoft IIS
& Java, Python, PHP, Node.js, Ruby, ASP.NET and Go\\
\parbox[t]{0.18\textwidth}{\heroku \\[-3pt] {\tiny\url{www.heroku.com}}}
& Jetty, Tornado, PHP built-in webserver, Mongrel, Thin, Hypnotoad, Mongoose, Yaws, Mochiweb
& Java, Python, PHP, Node.js, Ruby, Go, Perl, C, Erlang, Scala, and Clojure\\
\parbox[t]{0.18\textwidth}{\hpcloud \\[-3pt] {\tiny\url{www.hpcloud.com}}}
& Apache, Apache TomEE, Nginx
& Java, Python, PHP, Node.js, Ruby, Perl, Erlang, Scala, Clojure, ASP.NET \\
\parbox[t]{0.18\textwidth}{\openshift \\[-3pt] {\tiny\url{www.openshift.com}}}
& JBoss, Wildfly, Apache Tomcat, Zend Server, Vert.x
& Java, Python, PHP, Node.js, Ruby, Perl, Ceylon, and others\\
\bottomrule
\end{tabular}
}
\caption{Server and programming-language support in various PaaS clouds}
\label{tbl:survey}
\end{table*}

\section{Evaluation of \sysname's Memory Savings}
\label{sec:memory}

In this appendix we test the memory footprints induced by \sysname's
copy-on-access mechanism in comparison to the simpler alternative of
simply not sharing memory between containers at all.  (We caution the
reader that this simpler alternative addresses only \flushreload
attacks; it could not serve as a replacement for our \primeprobe
defense in \secref{sec:cacheabilityMgmt}.)  To measure the memory
savings that copy-on-access offers over disabling memory sharing
between containers, we measured the total unique physical memory pages
used across various numbers of webservers, each in its own container,
when running over (i) unmodified Linux, (ii) Linux without
cross-container memory sharing, and (iii) \sysname-enabled Linux.  We
used the system diagnosis tool \smem for memory accounting,
specifically by accumulating the PSS (proportional set size) field
output by \smem for each process, which reports the process' shared
memory pages divided by the number of processes sharing these pages,
plus its unshared memory pages.

\begin{figure}[h]
\centering
\subfigure[][Webservers idle]{
\includegraphics[width=0.85\columnwidth]{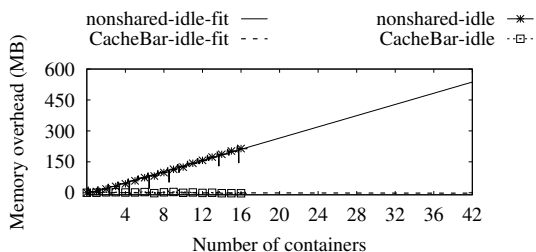}
\label{fig:mem:idle}
}
\subfigure[][25\% of webservers busy]{
\includegraphics[width=0.85\columnwidth]{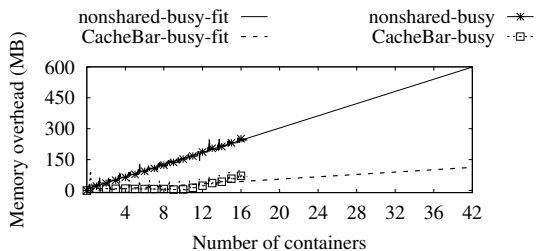}
\label{fig:mem:busy}
}
\caption{Memory overhead compared with unmodified Linux}
\label{fig:mem}
\end{figure}

For each platform, we incrementally grew the number of containers,
each containing a webserver, and left each webserver idle after
issuing to it a single request to confirm its functioning.  For each
number of containers, we measured the memory usage on the machine.
\figref{fig:mem:idle} shows the memory overhead of Linux without
cross-container sharing (``nonshared-idle'') and \sysname
(``\sysname-idle''), computed by subtracting the memory measured for
unmodified Linux from the memory measured for each of these systems.  We
grew the number of containers to 16 in each case, and then
extrapolated these measurements to larger numbers of containers using
best-fit lines (``nonshared-idle-fit'' and ``\sysname-idle-fit'').  As
can be seen in \figref{fig:mem:idle}, the overhead of \sysname is
virtually zero, whereas the overhead of Linux without cross-container
sharing is more substantial, even with negligible query load.

The memory overhead of \sysname does grow somewhat (relative to that
of unmodified Linux) when some of the servers are subjected to load.
\figref{fig:mem:busy} shows the same measures, but in an experiment in
which every fourth server was subjected to a slightly more active (but
still quite modest) load of four requests per second.  This was enough
to induce \sysname's copy-on-access mechanism to copy some memory
pages, resulting in a more noticeable increase in the memory usage of
the containers on \sysname-enabled Linux.  Again, however, the memory
overhead of \sysname was substantially less than of disabling
cross-container sharing altogether.  Moreover, even at maximum
throughput load for all servers (not shown), the \sysname overhead
approaches but does not exceed that of disabling cross-container
sharing.  As such, the copy-on-access mechanism strikes a better
balance between minimizing memory footprints and isolating containers
from side-channels than does simply disabling cross-container sharing.

\end{document}